\documentclass[traditabstract]{aa}

\usepackage{graphicx}
\usepackage{txfonts}
\usepackage{array}
\usepackage{amssymb}
\usepackage{natbib}
\usepackage{lscape}
\usepackage{rotating}
\usepackage{longtable}
\newcommand{\pa}{$\pi$\,Aqr}
\newcommand{\kms}{km\,s$^{-1}$}
\newcommand{\xmm}{{\sc{XMM}}\emph{-Newton}}

\newcommand{\sw}{{\em Swift}}
\begin{document}

\title{Surprises in the simultaneous X-ray and optical monitoring of \pa \thanks{Based on spectra obtained with the TIGRE telescope, located at La Luz observatory, Mexico (TIGRE is a collaboration of the Hamburger Sternwarte, the Universities of Hamburg, Guanajuato, and Li\`ege), as well as data collected with the Neil Gehrels \sw\ Observatory.}}

\author{Ya\"el~Naz\'e\inst{1}\thanks{F.R.S.-FNRS Senior Research Associate.}
\and Gregor~Rauw\inst{1} \and Myron Smith\inst{2}
}

\institute{Groupe d'Astrophysique des Hautes Energies, STAR, Universit\'e de Li\`ege, Quartier Agora (B5c, Institut d'Astrophysique et de G\'eophysique), All\'ee du 6 Ao\^ut 19c, B-4000 Sart Tilman, Li\`ege, Belgium\\
  \email{ynaze@uliege.be}
  \and National Optical Astronomy Observatory, 950 N Cherry Ave, Tucson, AZ 85721, USA
}

\authorrunning{Naz\'e et al.}
\titlerunning{\pa\ in X-rays and visible }
\abstract{To help constrain the origin of the peculiar X-ray emission of $\gamma$\,Cas stars, we conducted a simultaneous optical and X-ray monitoring of \pa\ in 2018. At that time, the star appeared optically bright and active, with a very strong H$\alpha$ emission. Our monitoring covers three 84\,d orbital cycles, allowing us to probe phase-locked variations as well as longer-term changes. In the new optical data, the radial velocity variations seem to span a smaller range than previously reported, which might indicate possible biases. The X-ray emission is variable, but without any obvious correlation with orbital phase or H$\alpha$ line strength. Furthermore, the average X-ray flux and the relative range of flux variations are similar to those recorded in previous data, although the latter data were taken when the star was less bright and its disk had nearly entirely disappeared. Only the local absorption component in the X-ray spectrum appears to have strengthened in the new data. This absence of large changes in X-ray properties despite dramatic disk changes appears at odds with previous observations of other $\gamma$\,Cas stars. It also constrains scenarios proposed to explain the $\gamma$\,Cas phenomenon. }
\maketitle

\section{Introduction}
Massive stars have decades ago been found to emit X-rays. This high-energy emission may be linked to accretion onto a compact object either from their stellar wind or through Roche-lobe overflow, when they are part of high-mass X-ray binaries. The ensuing X-ray emission is hard and extremely intense. On the other hand, shocks within the stellar wind or between two stellar winds in massive binaries also lead to the emission of X-rays, but this thermal emission is much softer ($kT\sim0.6-2.$\,keV) and dimmer ($\log[L_{\rm X}/L_{\rm BOL}]\sim-7$ to --6).

In between these two extremes are $\gamma$\,Cas analogs, named after their prototype, the central star of the Cassiopeia constellation. These stars exhibit a thermal spectrum with a high plasma temperature ($kT$=5--15\,keV) and a fluorescent iron line at 6.4\,keV in the iron complex. Their (variable) brightness is intermediate between ``normal'' OB-stars and X-ray binaries ($\log[L_{\rm X}/L_{\rm BOL}]$ between --6 and --4). Currently, about 20 such objects are known \citep[see][]{smi16,naz18}.

The origin of the peculiar X-ray emission of $\gamma$\,Cas stars remains debated. Two main classes of scenarios have been proposed. On the one hand, X-rays could be emitted by an accreting white dwarf \citep{mur86} or by an accreting neutron star in the propeller regime \citep{pos17}. On the other hand, X-rays could be emitted by star-disk interactions occurring through localized magnetic fields in the disk, which could be generated by disk instabilities \citep{smi98,smi99}, and in the star, possibly arising from a subsurface equatorial convective zone \citep{can09,mot15}. In both cases, changes in the disk are expected to eventually affect the X-ray emission, possibly after some delay, depending on the model under consideration.

\begin{table}
\centering
\caption{ Properties of \pa\  as reported by \citet{bjo02}.  }
\label{paprop}
\setlength{\tabcolsep}{3.3pt}
\begin{tabular}{cc}
\hline\hline
Parameter & Value \\
\hline
$P_{orb}$ & 84.1\,d\\
$T_0$    & 2\,450\,275.5\\
$q=M_2/M_1$ & 0.16\\
$a\,\sin(i)$ & 0.96\,AU \\
$M_1$ & 11$\pm$1.5\,M$_{\odot}$ or 15$\pm$3\,M$_{\odot}$\\
$R_1$ & 6.1$\pm$2.5\,R$_{\odot}$\\
$T_{eff}$ & 25$\pm$2\,kK\\
$\log(g)$ & 3.9$\pm$0.1\\
$v\,\sin(i)$ & 250\,\kms \\
$i$ & 70$^{\circ}$ (50--75$^{\circ}$)\\
\hline      
\end{tabular}
\\
\tablefoot{ Average values are shown for $P_{orb}$ and $T_0$. The two mass values come from a comparison of evolutionary tracks and dynamical analysis.}
\end{table}

Monitoring therefore is an essential tool for understanding the $\gamma$\,Cas phenomenon, and we therefore decided to undertake a simultaneous X-ray and optical monitoring of \pa\ in 2018. This recently discovered \citep{naz17} example of a $\gamma$\,Cas star is a known binary with a period of 84.1\,d \citep[see Table \ref{paprop}]{bjo02}. If the peculiar X-ray emission of $\gamma$\,Cas stars is linked to accretion, the low-mass companion detected by \citet{bjo02} would have to be the compact accretor (although these authors favored a non-degenerate nature). There is no room in the system for a third object to orbit close to (and to accrete from) the disk in a stable way.

Our monitoring fully covered three orbital cycles of \pa\  in order to ascertain the presence of phase-locked changes of the X-ray emission. These variations could occur because the orbital inclination of the system is rather high (70$^{\circ}$) and the companion was known to affect the disk geometry \citep{zha13}. Comparing the new data to older X-ray detections also allows us to study the longer-term variations, if any. This article reports the results of this variability study. After a brief summary of the observations and their reduction (Sect. 2), the behavior of \pa\ in the optical and X-ray domains is examined (Sect. 3 and 4, respectively) before we draw a short conclusion (Sect. 5).

\begin{table}
\centering
\caption{Journal of the TIGRE observations of \pa.  }
\label{journalHRT}
\setlength{\tabcolsep}{3.3pt}
\begin{tabular}{ccccccc}
\hline\hline
HJD & Exp. Time & $-EW$(H$\alpha$) & \multicolumn{3}{c}{$RV$(H$\alpha$, in \kms)} &\# \\
-2\,458\,000. & (s) & (\AA) & $M_1$ & mirror & 2G &\\
\hline
 230.766 &  709 & 21.96 &  -5.3 &  -4.5 &  -4.0 & 2  \\   
 242.755 &  872 & 21.81 &  -8.2 &  -8.1 &  -7.0 &    \\    
 255.734 &  875 & 21.79 & -12.2 & -14.2 & -13.4 & 5  \\   
 266.721 &  686 & 22.16 & -16.5 & -19.8 & -19.2 & 7  \\    
 274.717 &  717 & 22.62 & -15.5 & -20.2 & -20.0 & 8  \\    
 301.669 & 1100 & 22.16 &  -8.3 &  -8.0 &  -8.9 &    \\    
 309.686 & 1420 & 21.93 &  -6.6 &  -4.6 &  -5.7 &13  \\    
 319.597 &  909 & 22.25 &  -8.1 &  -6.0 &  -7.7 &14  \\    
 325.630 &  906 & 22.71 &  -9.5 &  -7.5 &  -8.9 &16-7\\    
 333.622 & 1865 & 22.64 & -13.9 & -12.9 & -13.6 &18  \\   
 347.635 &  898 & 23.27 & -17.6 & -17.7 & -17.7 &20  \\   
 355.554 &  698 & 23.94 & -18.4 & -18.6 & -18.3 &20  \\   
 366.490 &  747 & 23.66 & -17.4 & -18.0 & -17.5 &22  \\   
 366.546 &  849 & 23.37 & -17.7 & -17.6 & -17.2 &22  \\   
 382.571 & 1343 & 23.76 &  -7.8 &  -6.8 &  -6.2 &24  \\   
 391.467 & 1664 & 24.06 &  -6.8 &  -4.9 &  -4.9 &26-7\\   
 400.479 &  757 & 24.91 &  -6.1 &  -4.1 &  -4.2 &28  \\   
 446.320 &  704 & 24.37 & -14.0 & -18.2 & -17.6 &36  \\   
 458.293 &  796 & 24.23 &  -7.7 & -12.0 & -11.3 &39  \\   
 470.321 &  635 & 24.73 &  -2.8 &  -5.5 &  -4.9 &39  \\   
 479.313 &  940 & 24.57 &  -1.9 &  -3.8 &  -3.6 &    \\    
\hline      
\end{tabular}
\\
\tablefoot{ $EW$ and $M_1$ correspond to zeroth- and first-order moments (see Section 3) evaluated between --540\,\kms\ and +540\,\kms, ``mirror'' indicates results from the mirror method for determining the radial velocity, and ``2G'' those from the two-Gaussian correlation method (see Section 3.1 for details). A rest wavelength of 6562.85\,\AA\ was used for the H$\alpha$ line. The last column lists the end of the \sw\ ObsID used to correlate X-ray and optical data (see Section 4.2 for details).}
\end{table}

\begin{table*}
\centering
\caption{X-ray properties of \pa\ derived from \sw\ observations.  }
\label{journal}
\setlength{\tabcolsep}{3.3pt}
\begin{tabular}{ccc|cc|ccccc}
\hline\hline
ObsID & HJD           &  Exp. Time & Ct rate in T  & $HR=H/S$       & $N_{\rm H}$    & $norm$         & $F^{\rm obs}_{\rm X}$ & $F^{\rm abs-cor}_{\rm X}$ & $\chi_{\nu}^2$ (dof) \\
      & -2\,458\,000. & (s) & (cts\,s$^{-1}$) &       &  (10$^{22}$\,cm$^{-2}$) & ($10^{-3}$\,cm$^{-5}$) &  \multicolumn{2}{c}{($10^{-11}$\,erg\,cm$^{-2}$\,s$^{-1}$)} & \\
\hline
00010659001 & 223.678 & 1281 & 0.224$\pm$0.017 & 0.91$\pm$0.14 & 0.39$\pm$0.12 & 8.04$\pm$0.73 & 1.27$\pm$0.11 & 1.48$\pm$0.14 & 1.30(28)\\
00010659002 & 232.494 & 1426 & 0.288$\pm$0.030 & 0.76$\pm$0.16 & 0.11$\pm$0.10 & 6.13$\pm$0.65 & 1.05$\pm$0.08 & 1.13$\pm$0.12 & 1.40(30)\\
00010659003 & 236.481 &  980 & 0.240$\pm$0.021 & 0.57$\pm$0.10 & 0.18$\pm$0.11 & 7.10$\pm$0.78 & 1.19$\pm$0.11 & 1.31$\pm$0.15 & 0.60(18)\\
00010659005 & 255.425 &  968 & 0.301$\pm$0.021 & 0.68$\pm$0.10 & 0.20$\pm$0.09 & 9.72$\pm$0.85 & 1.62$\pm$0.12 & 1.79$\pm$0.14 & 0.76(27)\\
00010659006 & 261.000 & 1003 & 0.211$\pm$0.019 & 0.89$\pm$0.16 & 0.51$\pm$0.13 & 8.18$\pm$0.82 & 1.26$\pm$0.10 & 1.50$\pm$0.09 & 1.23(17)\\
00010659007 & 267.237 & 1018 & 0.297$\pm$0.019 & 0.94$\pm$0.12 & 0.45$\pm$0.11 &10.87$\pm$0.97 & 1.69$\pm$0.11 & 2.00$\pm$0.18 & 0.73(26)\\
00010659008 & 272.692 & 1008 &                 &               & 0.90$\pm$0.27 & 6.97$\pm$0.94 & 1.01$\pm$0.11 & 1.28$\pm$0.17 & 0.69(15)\\
00010659009 & 278.858 &  888 & 0.208$\pm$0.017 & 4.19$\pm$1.01 & 3.15$\pm$0.64 &14.95$\pm$2.00 & 1.76$\pm$0.20 & 2.75$\pm$0.37 & 1.06(16)\\
00010659010 & 285.368 & 1013 & 0.308$\pm$0.020 & 1.53$\pm$0.21 & 1.76$\pm$0.33 &15.07$\pm$1.58 & 1.97$\pm$0.16 & 2.77$\pm$0.27 & 1.22(27)\\
00010659011 & 291.413 &  389 & 0.204$\pm$0.027 & 0.95$\pm$0.25 & 0.76$\pm$0.38 & 7.24$\pm$1.45 & 1.07$\pm$0.18 & 1.33$\pm$0.28 & 1.53(5) \\
00010659013 & 309.148 &  824 & 0.339$\pm$0.025 & 0.66$\pm$0.09 & 0.15$\pm$0.09 & 9.69$\pm$0.98 & 1.64$\pm$0.14 & 1.78$\pm$0.18 & 1.15(21)\\
00010659014 & 317.543 &  920 & 0.236$\pm$0.018 & 1.22$\pm$0.19 & 0.70$\pm$0.19 &10.02$\pm$1.09 & 1.49$\pm$0.14 & 1.84$\pm$0.21 & 0.63(20)\\
00010659016 & 324.685 & 1131 &                 &               & 1.91$\pm$0.49 & 7.92$\pm$1.26 & 1.02$\pm$0.15 & 1.46$\pm$0.24 & 1.53(10)\\
00010659017 & 326.626 &  923 & 0.275$\pm$0.030 & 2.07$\pm$0.44 & 1.69$\pm$0.44 &13.72$\pm$2.10 & 1.81$\pm$0.21 & 2.53$\pm$0.40 & 1.22(16)\\
00010659018 & 332.970 &  894 & 0.218$\pm$0.031 & 0.90$\pm$0.21 & 0.43$\pm$0.20 &10.65$\pm$1.48 & 1.67$\pm$0.20 & 1.96$\pm$0.21 & 0.87(13)\\
00010659019 & 338.971 &  820 & 0.230$\pm$0.032 & 1.13$\pm$0.32 & 0.83$\pm$0.26 & 9.99$\pm$1.48 & 1.45$\pm$0.19 & 1.84$\pm$0.28 & 0.52(9) \\
00010659020 & 351.253 & 1008 & 0.285$\pm$0.020 & 1.21$\pm$0.18 & 0.63$\pm$0.15 &11.13$\pm$1.04 & 1.68$\pm$0.13 & 2.05$\pm$0.19 & 0.74(27)\\
00010659022 & 362.741 &  923 & 0.269$\pm$0.020 & 0.95$\pm$0.14 & 0.58$\pm$0.14 & 9.61$\pm$0.92 & 1.46$\pm$0.12 & 1.77$\pm$0.17 & 0.87(21)\\
00010659024 & 386.429 &  879 & 0.333$\pm$0.023 & 1.87$\pm$0.29 & 1.01$\pm$0.21 &14.91$\pm$1.51 & 2.12$\pm$0.16 & 2.74$\pm$0.28 & 0.97(23)\\
00010659026 & 390.210 &  960 & 0.195$\pm$0.019 & 1.43$\pm$0.30 & 0.67$\pm$0.17 & 9.60$\pm$0.98 & 1.43$\pm$0.12 & 1.77$\pm$0.18 & 0.33(19)\\
00010659027 & 392.907 & 1013 & 0.328$\pm$0.022 & 1.35$\pm$0.18 & 0.68$\pm$0.14 &13.58$\pm$1.18 & 2.03$\pm$0.14 & 2.50$\pm$0.37 & 0.73(30)\\
00010659028 & 399.079 &  922 & 0.325$\pm$0.027 & 1.14$\pm$0.19 & 0.41$\pm$0.11 &12.49$\pm$1.10 & 1.96$\pm$0.13 & 2.30$\pm$0.20 & 0.86(29)\\
00010659029 & 405.390 & 1038 & 0.191$\pm$0.018 & 1.24$\pm$0.25 & 0.80$\pm$0.19 & 9.24$\pm$0.97 & 1.35$\pm$0.12 & 1.70$\pm$0.18 & 0.68(17)\\
00010659031 & 416.608 & 1013 & 0.280$\pm$0.022 & 1.27$\pm$0.21 & 0.39$\pm$0.12 &10.36$\pm$0.97 & 1.64$\pm$0.14 & 1.90$\pm$0.18 & 1.22(25)\\
00010659032 & 422.717 &  995 & 0.334$\pm$0.023 & 0.90$\pm$0.12 & 0.19$\pm$0.07 &10.85$\pm$0.85 & 1.81$\pm$0.13 & 1.99$\pm$0.16 & 0.66(29)\\
00010659033 & 429.427 &  978 & 0.229$\pm$0.020 & 1.16$\pm$0.20 & 0.46$\pm$0.24 & 8.33$\pm$1.19 & 1.30$\pm$0.12 & 1.54$\pm$0.21 & 1.19(22)\\
00010659034 & 435.073 & 1098 & 0.152$\pm$0.014 & 2.65$\pm$0.60 & 1.72$\pm$0.39 & 9.00$\pm$1.21 & 1.18$\pm$0.12 & 1.66$\pm$0.22 & 0.58(15)\\
00010659035 & 441.179 &  978 & 0.203$\pm$0.019 & 1.05$\pm$0.20 & 0.31$\pm$0.12 & 7.23$\pm$0.78 & 1.16$\pm$0.11 & 1.33$\pm$0.14 & 0.99(17)\\
00010659036 & 446.889 &  812 & 0.246$\pm$0.022 & 0.87$\pm$0.15 & 0.52$\pm$0.15 &10.38$\pm$1.11 & 1.60$\pm$0.15 & 1.91$\pm$0.15 & 0.69(18)\\
00010659039 & 465.012 &  983 &                 &               & 0.63$\pm$0.35 & 4.76$\pm$0.94 & 0.72$\pm$0.13 & 0.88$\pm$0.14 & 1.73(7) \\
\hline      
\end{tabular}
\\
\tablefoot{For count rates, total (T), soft (S), and hard (H) bands are defined as 0.5--10.\,keV, 0.5--2.\,keV, and 2.--10.\,keV, respectively. For spectral fits, a model $tbabs\times phabs\times apec$ was used, with the first absorption fixed to the interstellar value of $3.6\times10^{20}$\,cm$^{-2}$ and the temperature fixed to 14.8\,keV (see text for details). Errors (found using the ``error'' command for the spectral parameters and the absorption-corrected fluxes, or the ``flux err'' command for the observed fluxes) correspond to 1$\sigma$; when the errors were asymmetric, the highest value is provided here. Fluxes were determined in the 0.5--10.0\,keV band; absorption-corrected fluxes, found using ``cflux'' in front of the thermal component, are corrected for the full (i.e., interstellar+local) absorbing column. }
\end{table*}

\section{Observations and data reduction} 

\subsection{Optical spectroscopy}
We monitored three orbits of \pa\ in April-December 2018 with the fully robotic 1.2\,m TIGRE telescope \citep{Schmitt} installed at La Luz Observatory near Guanajuato (Mexico). The telescope is equipped with the refurbished HEROS echelle spectrograph covering the wavelength domain from 3800 to 8800\,\AA, with a small 100\,\AA\ wide gap near 5800\,\AA. The resolving power is about 20\,000. The data reduction was performed with the dedicated TIGRE/HEROS reduction pipeline \citep{Mittag}. Absorption by telluric lines was corrected within IRAF using the template of \citet{hin00} around the He\,{\sc i}\,$\lambda$\,5876\AA, H$\alpha$, O, and Fe lines in 7400--7850\AA, as well as the Paschen lines at 8350--8800\AA. As a last step, the high-resolution spectra were normalized over limited wavelength windows using splines of low order. The journal of the observations is provided in Table \ref{journalHRT}.

\begin{figure*}
  \begin{center}
\includegraphics[width=6cm]{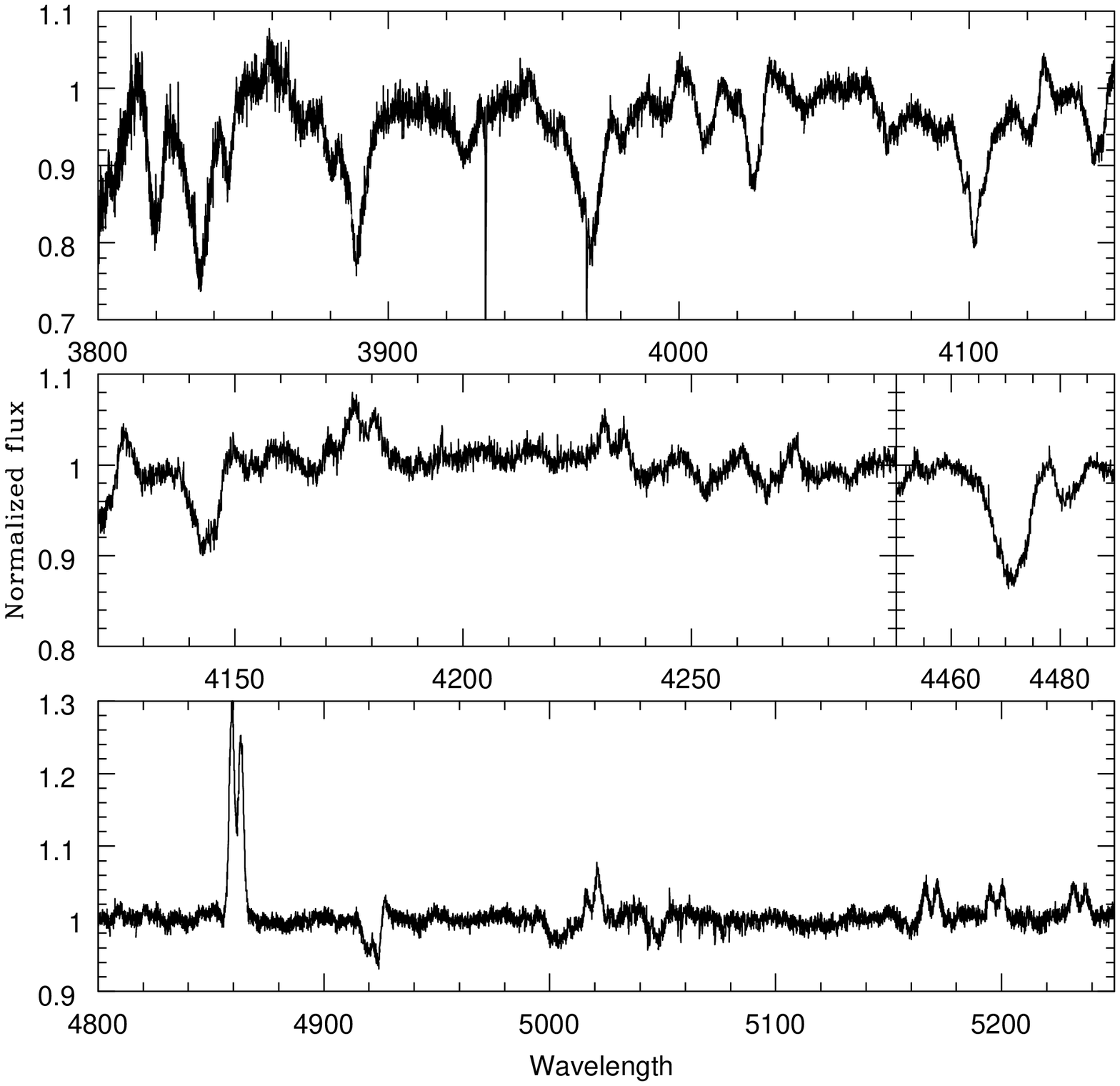}
\includegraphics[width=6cm]{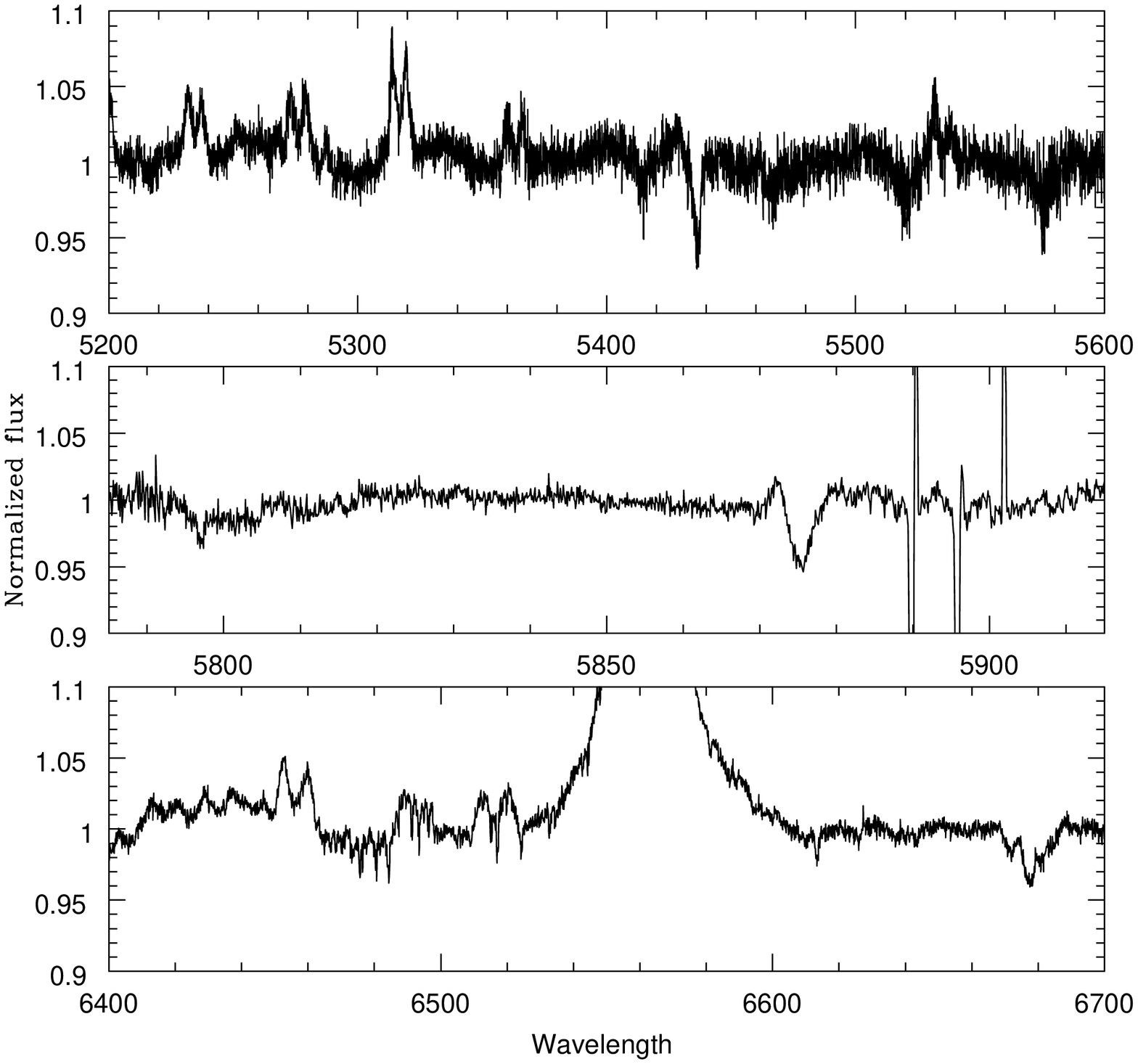}
\includegraphics[width=6cm]{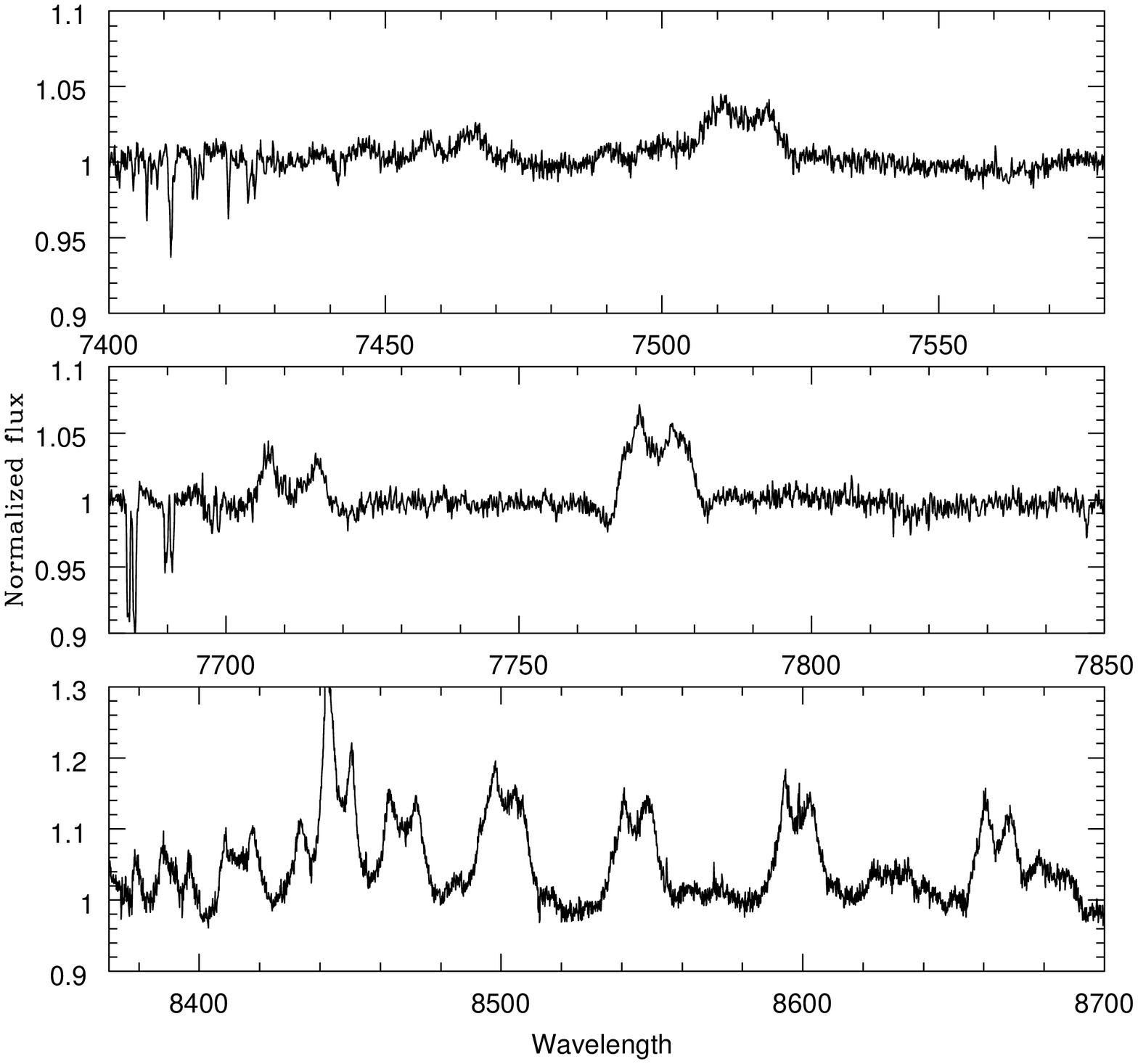}
  \end{center}
\caption{Spectrum of \pa\ taken by TIGRE on 10 July 2018, after correction of the telluric lines. }
\label{spec}
\end{figure*}

\subsection{X-ray data}
Simultaneously with the optical campaign, \pa\ was observed 35 times by the Neil Gehrels \sw\ Observatory in April-December 2018. One of these observations (ObsID=00010659015) was stopped before pointing had settled, the source appeared on the edge of the detector in another observation (ObsID=00010659030), and it fell onto bad pixels in a third observation (ObsID=00010659037), making these three observations unusable. Two additional observations (ObsID=00010659012 and 21) lasted less than 200s, hence were too short to provide a detectable signal for the target. Table \ref{journal} provides the journal of the remaining 30 observations, which have a typical duration of 1\,ks.

Because \pa\ is a bright source at optical and UV wavelengths, the XRT data were taken in Windowed Timing mode to avoid optical loading. To eliminate any remaining contamination, we used only the best-quality ($grade=0$) events. The UK online tool\footnote{http://www.swift.ac.uk/user\_objects/ \citep{eva09}} was used to extract spectra for each observation as well as corrected (full point spread function, PSF) count rates in the 0.5-10.\,keV energy band. No reliable count rates could be produced online for three observations (ObsID=00010659008, 16, and 39), however, hence they are missing from Table~\ref{journal}.

In parallel, we also processed the data locally using the XRT pipeline of HEASOFT v6.22.1 with calibrations v20170501. The source spectra were extracted within
a circular aperture centered on the Simbad coordinates of the source. Following recommendations of the XRT team, a small radius of 20\,px was used to minimize the background contribution; in a few cases, when the source appeared closer to the edge, it was even further reduced to 10\,px. For the background estimate, we again followed the XRT team recommendations and used the surrounding annular region with radii 70 and 130\,px. Because Windowed Timing data from a single snapshot are compressed in a single row, the spectral scaling parameter ($BACKSCAL$ keyword in header) had to be modified from the areas of the circular regions to their diameters\footnote{http://www.swift.ac.uk/analysis/xrt/backscal.php}, that is, to a value of 40 for the source and 59 for the background. The adequate RMF matrix from the calibration database was used, and specific ARF response matrices were calculated for each dataset using {\it xrtmkarf}, considering the associated exposure map. Both local and online spectra were finally binned using {\it grppha} to reach at least 10 cts per bin.

\section{Optical and near-IR lines of \pa}

The optical and near-infrared (NIR) spectrum of \pa\ from our data is shown in Fig. \ref{spec}. It is dominated by strong emissions of hydrogen (Balmer H$\alpha$ and H$\beta$, as well as Paschen lines). Additional emissions, linked to Fe\,{\sc ii} but also to O\,{\sc i} and O\,{\sc iii}, are detected. Absorption lines include He\,{\sc i} lines and several interstellar lines (Na\,{\sc i}, Ca\,{\sc ii}, and diffuse interstellar bands; DIBs). Some lines, such as other Balmer lines and He\,{\sc i}\,$\lambda\lambda$4471,5876\AA, display a mixed absorption and emission content, however.

To characterize the lines, we calculated moments after subtracting 1 from the normalized (see Sect. 2.1) spectra: $M_0=\sum (F_i-1)$, $M_1=\sum (F_i-1)\times v_i /\sum (F_i-1)$, and $M_2=\sum (F_i-1)\times (v_i-M_1)^2 /\sum (F_i-1),$ where $v_i$ is the radial velocity and $F_i$ the normalized flux. They provide the equivalent width\footnote{after multiplication by minus the wavelength step, to have the usual definition of positive $EWs$ for absorptions and negative $EWs$ for emissions.} $EW$, the line centroid, and the square of the line width, respectively. For each line, we limited the calculation to a wavelength interval covering most of the profile while avoiding nearby lines or the continuum regions. While the errors on moments can be calculated using error propagation, the interstellar lines provide a check and typical values. The average moments of the interstellar Na\,{\sc i}\,$\lambda$\,5895.924\AA\ line, estimated between --25 and 5\,\kms, are $M_0=0.214\pm0.008$\AA, $M_1=-9.2\pm0.5$\,\kms, and a width $\sqrt{M_2}=6.8\pm0.3$\,\kms\ (the errors here correspond to dispersions around the means). In the same velocity intervals, the values for the interstellar Ca\,{\sc ii}\,$\lambda$\,3933.663\AA\ and Ca\,{\sc ii}\,$\lambda$\,3968.468\AA\ are $M_0=0.095\pm0.007$\AA, $M_1=-10.7\pm0.4$\,\kms, $\sqrt{M_2}=7.2\pm0.4$\,\kms, and $M_0=0.109\pm0.008$\AA, $M_1=-10.3\pm0.5$\,\kms, $\sqrt{M_2}=8.0\pm0.6$\,\kms, respectively. The $<$1\kms\ uncertainty in velocity is typical of the wavelength accuracy of TIGRE data (M. Mittag, private communication). Finally, the values for the DIB at 6613.62\AA, estimated between --50 and 60\,\kms, are $M_0=0.032\pm0.005$\AA, $M_1=-1.7\pm2.5$\,\kms, with a width $\sqrt{M_2}=26.2\pm1.7$\,\kms; the uncertainties are larger because the line is much weaker and broader. 

\begin{figure*}
  \begin{center}
\includegraphics[width=6cm]{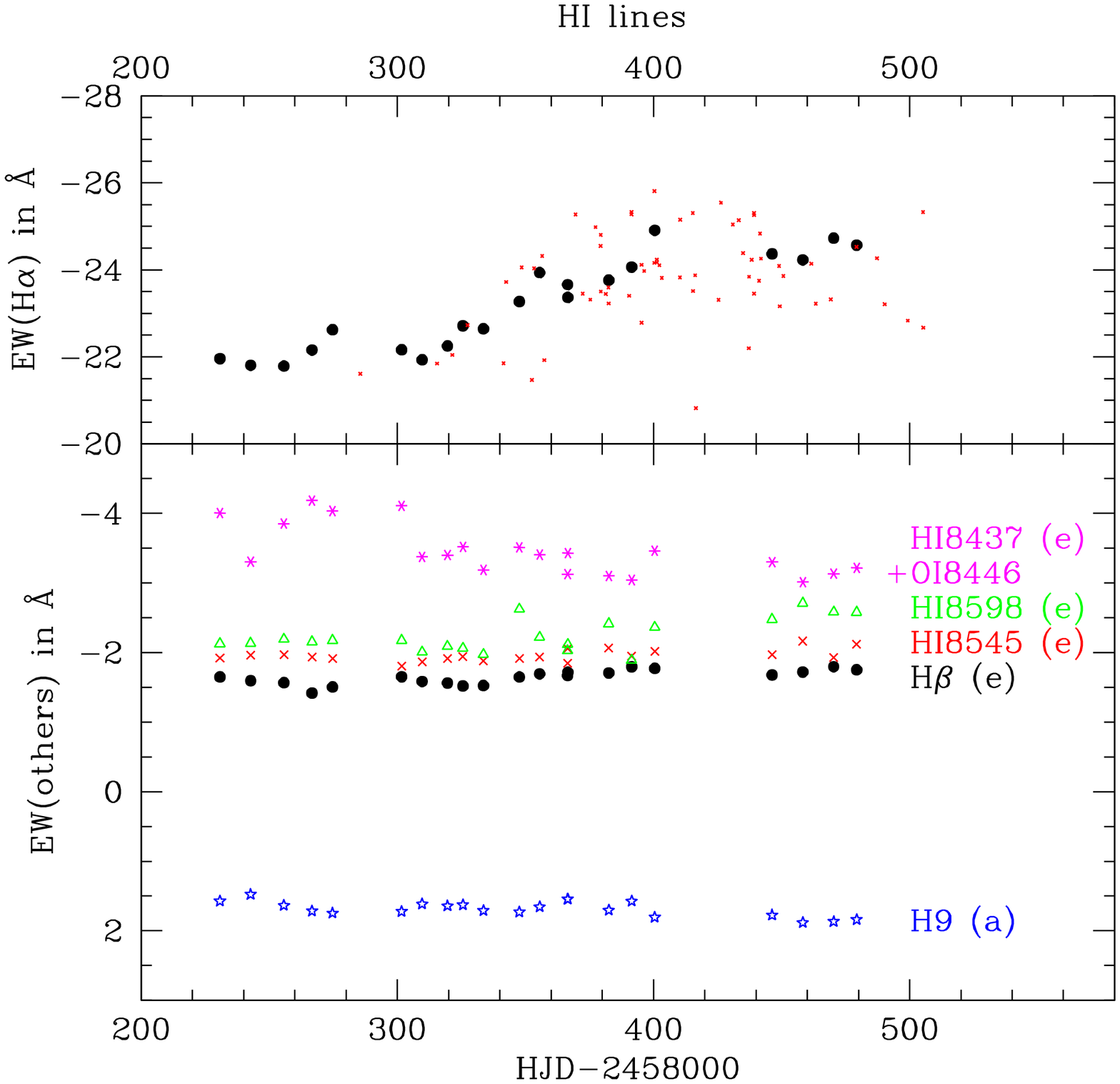}
\includegraphics[width=6cm]{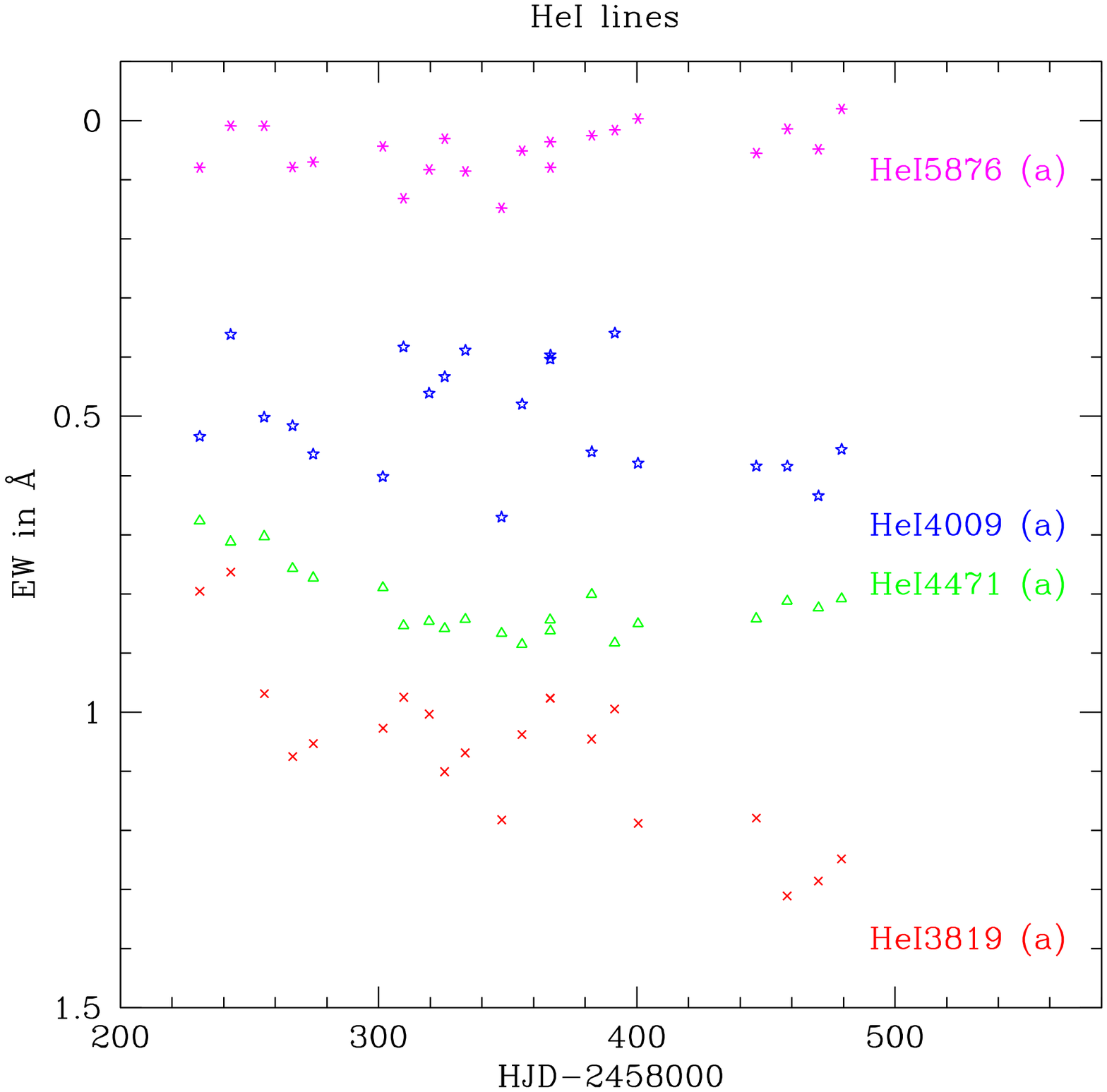}
\includegraphics[width=6cm]{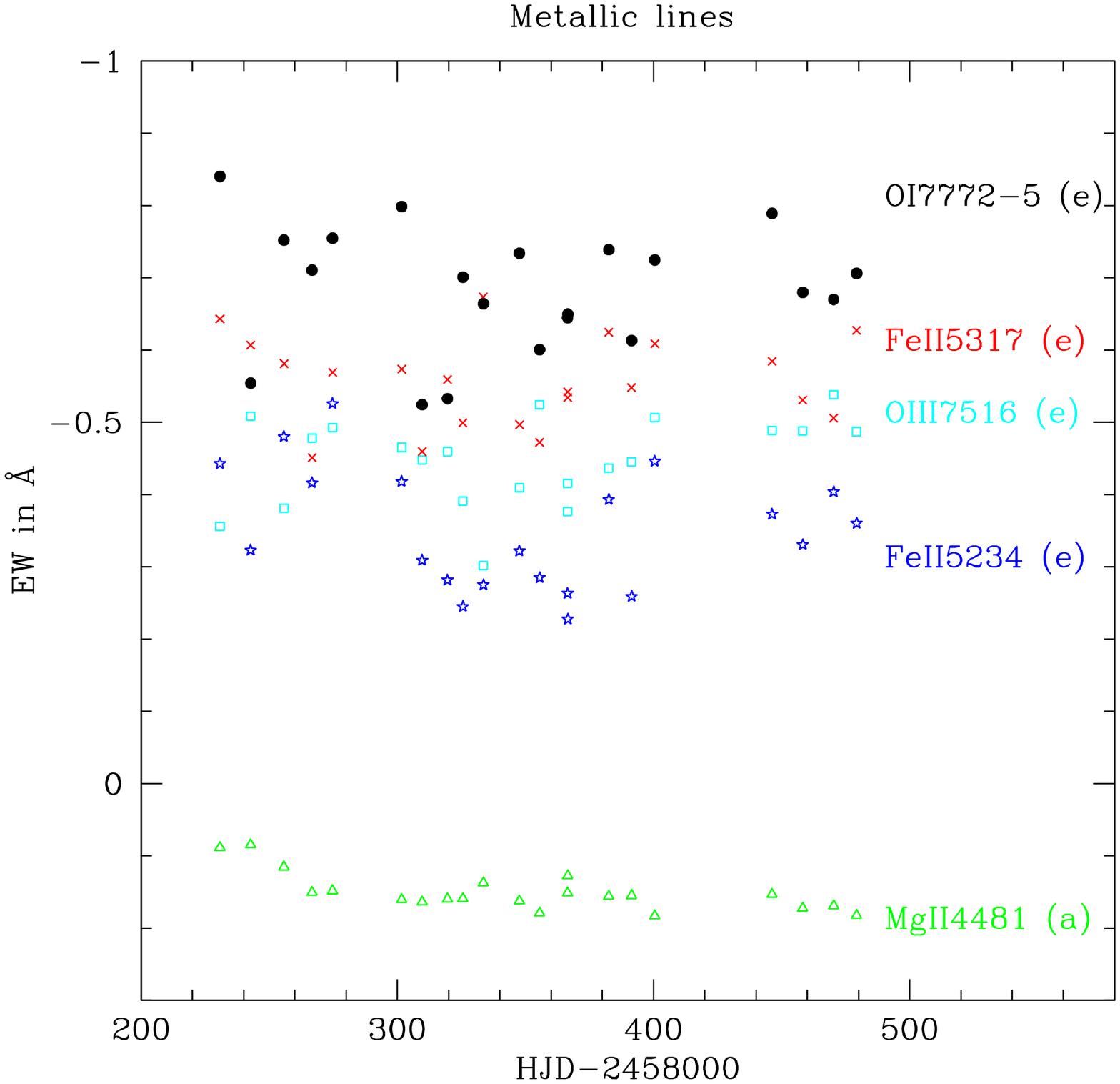}
  \end{center}
\caption{Evolution with time of zeroth-order moments ($EWs$) for hydrogen, He\,{\sc i}, and metallic lines. In these panels, higher emissions (absorptions) are toward the top (bottom). For H$\alpha$, the values derived from amateur data reported by \citet{naz19} are also shown as small red crosses. }
\label{ew}
\end{figure*}

\begin{figure}
  \begin{center}
\includegraphics[width=8.5cm]{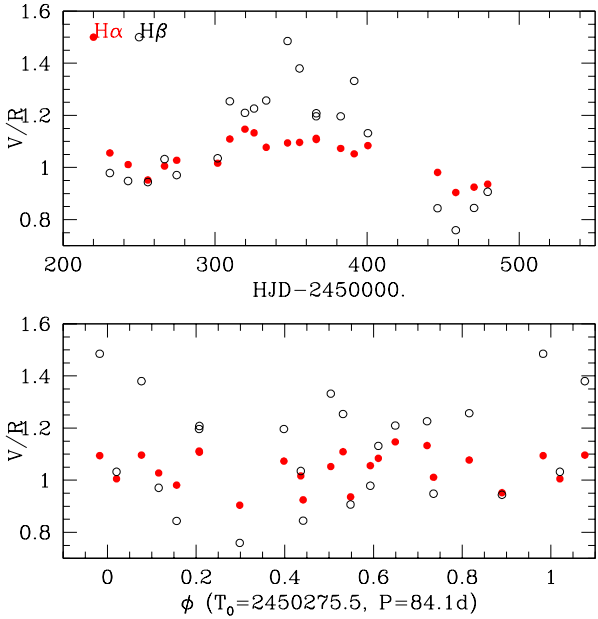}
  \end{center}
\caption{Evolution with time or phase of the $V/R$  for the H$\alpha$ and H$\beta$ lines. }
\label{ew2}
\end{figure}

Throughout our campaign, the H$\alpha$ emission strengthened, with $EWs$ evolving from --22 to --25\AA\ (Fig \ref{ew}, Table \ref{journalHRT}). Other hydrogen lines follow that behavior (i.e., more emission or less absorption in more recent times), except for H9, whose absorption increased, and the blends H\,{\sc i}\,$\lambda$8437\AA+O\,{\sc i}\,$\lambda$8446\AA\ and H\,{\sc i}\,$\lambda$8502\AA,\ whose emissions decreased. He\,{\sc i} and metallic lines displayed small changes in line strength but without a clear trend with time, except for He\,{\sc i}\,$\lambda$3819,4471\AA\ and Mg\,{\sc ii}\,$\lambda$4481\AA, whose absorptions increased. 

\begin{figure}
  \begin{center}
\includegraphics[width=8.5cm]{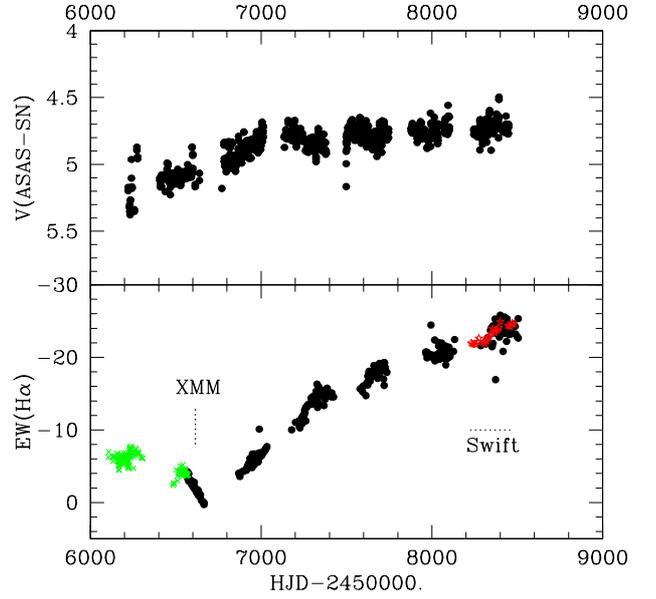}
  \end{center}
\caption{Evolution with time of $EW(H\alpha)$ (red stars from this paper, black dots from \citealt{naz19}, green crosses from \citealt{zha13}) and V magnitude from ASAS-SN (``bb'' camera only). Two dotted lines indicate the time of pointed X-ray observations (see Sect. 4).}
\label{photom}
\end{figure}

Finally, we note that the amplitude of the violet peak with respect to the red peak evolves with time, following a similar behavior in H$\alpha$ and H$\beta$ emissions (Fig \ref{ew2}). However, there is no clear correlation with orbital phase, as has been reported for recent years by \citet{naz19}. Moreover, \pa\ seems to have recently brightened in the optical domain ($V$ band) as its emission strengthened (Fig. \ref{photom}). A similar correlation was detected for $\gamma$\,Cas \citep{smi12}, but the opposite trend was found for HD\,45314 \citep{rau18}. These positive and negative correlations have been theoretically explained by \citet{sig13} by way of differences in disk inclination (higher inclinations leading to inverse correlations) but also in disk scale heights. 

\subsection{Radial velocities}

In addition to the first-order moment of the full profile, we evaluated the radial velocity ($RV$) of H$\alpha$ using two other techniques focusing on the line wings that were previously applied to the case of $\gamma$\,Cas. The first technique compares the line profile and its mirror (after reversing velocities) for different shifts, searching for the least difference between them \citep[and references therein]{nem12}. Only the wings between approximately 20\% and 60\% of the maximum amplitude were considered in this calculation for $\gamma$\,Cas. This corresponds to normalized fluxes of 1.4 and 2.2 for \pa\ because the profile peaks around a normalized flux of 3 in 2018. Shifts from --50\,\kms\ to +50\,\kms\ were examined, with steps of 0.5\,\kms. A parabolic fitting of the $\chi^2$ values was then made to find the final best shift, which corresponds to $-RV$. The second technique correlates the line profile to the two-Gaussians function $G(v)=exp[-(v-a)^2/2\sigma^2]-exp[-(v+a)^2/2\sigma^2]$ \citep[based on \citealt{sha86}]{smi12}. The $RV$ is then found when the correlation reaches zero. The Gaussian center $a$ and width $\sigma$ were chosen to be 245\,\kms\ and 20\,\kms, respectively, to allow us to derive the velocity near the half-maximum of the line (i.e., it is a bisector value).

\begin{figure}
  \begin{center}
\includegraphics[width=9cm, bb=30 400 550 660,clip]{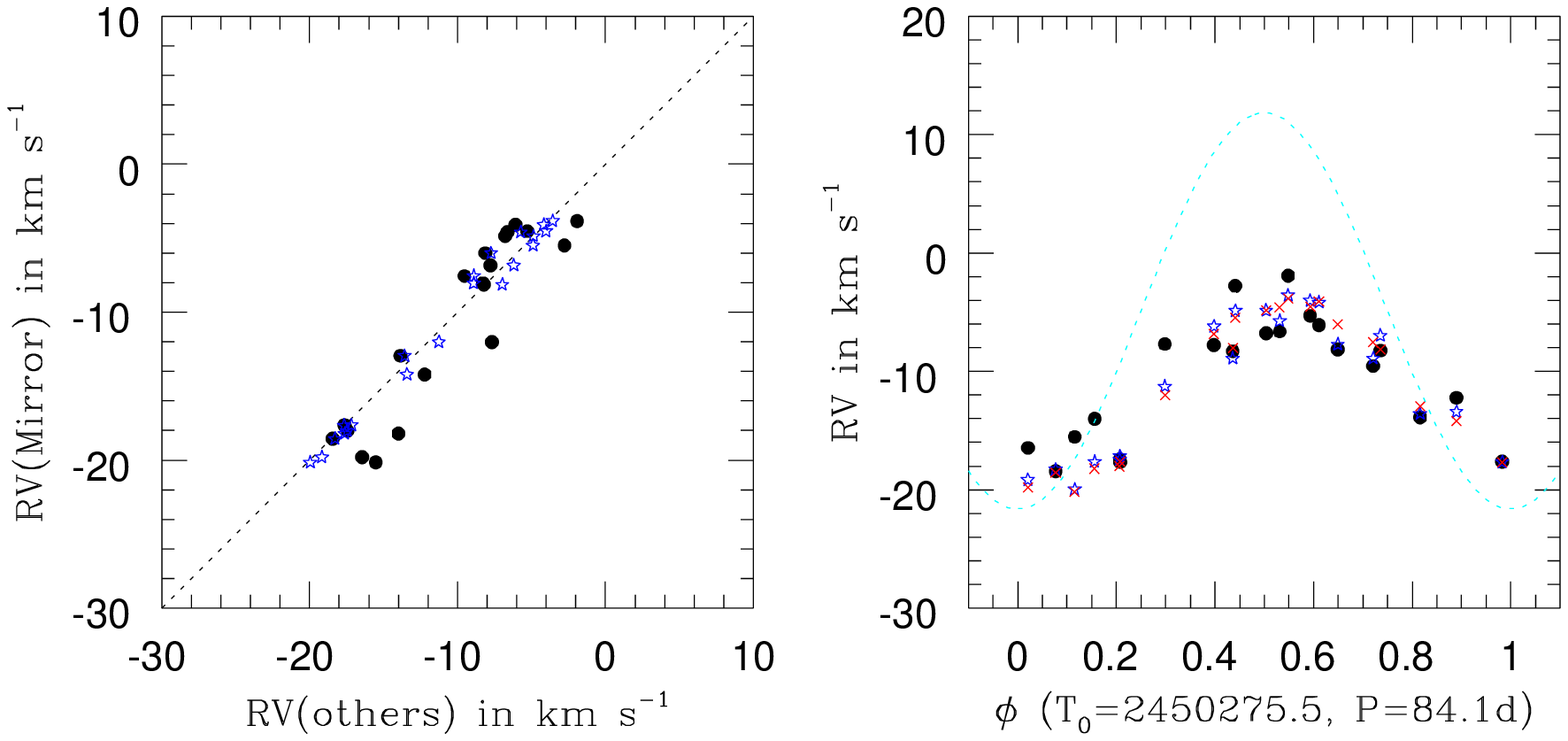}
  \end{center}
\caption{{\it Left:} Comparison between H$\alpha$ velocities derived through three different methods, with the velocities derived by the mirror method as ordinates and those derived from moments (black dots) and two-Gaussian correlation (blue stars) as abscissae. The dotted diagonal indicates equality. {\it Right:} Derived velocities phased with the average ephemeris of \citet{bjo02}. Symbols are black dots for the centroid derived from moments, blue stars for velocities derived from the two-Gaussian correlation, and red crosses for velocities derived by the mirror method. The dotted cyan line provides the orbital solution of \citet{bjo02}. }
\label{3rv}
\end{figure}

The left panel of Figure \ref{3rv} compares the velocities derived for the H$\alpha$ line by these three different techniques (see Table \ref{journalHRT} for values), revealing a good agreement between them. The dispersion for first-order moments appears larger, but this is to be expected because they probe the entire line profile, including the most strongly varying central regions. The right panel of the same figure displays these velocities folded with the mean ephemeris of \citet{bjo02}. It also compares them to the orbital solution of \citet{bjo02} for the primary star. We recall that this solution was derived by using the mirror method on the {\it \textup{absorption}} component (photosphere of the primary) because the disk had nearly completely disappeared at the time of their observations, while our values correspond to the velocities of the {\it \textup{emission}} component (disk of the primary) because the disk now strongly dominates the H$\alpha$ line.

Two conclusions can be drawn. First, there seems to be no phase shift, showing that the Bjorkman ephemeris remains accurate to this day. Second, a mismatch in amplitude is clearly detected: \citet{bjo02} inferred a semi-amplitude $K_1$ of 16.7\,\kms\ for the H$\alpha$ absorption observed at that time, while the semi-amplitude recorded for the H$\alpha$ emission in 2018 only amounts to half this value. At first, this decrease might be attributed to the blend with the companion line and/or the imperfect symmetry of the disk. Because the disk appears slightly brighter on the companion side (see section 3.2 and \citealt{naz19}), the derived velocities could be biased toward the companion velocity. A blend would have a similar effect. This would imply that velocities derived from the Be disk emissions may not be fully representative of the Be star's true velocity, although emissions such as H$\alpha$ are commonly used to this aim (see, e.g., \citealt{nem12} or \citealt{smi12} for $\gamma$\,Cas).

\begin{figure}
  \begin{center}
\includegraphics[width=8.5cm, bb=30 150 550 400]{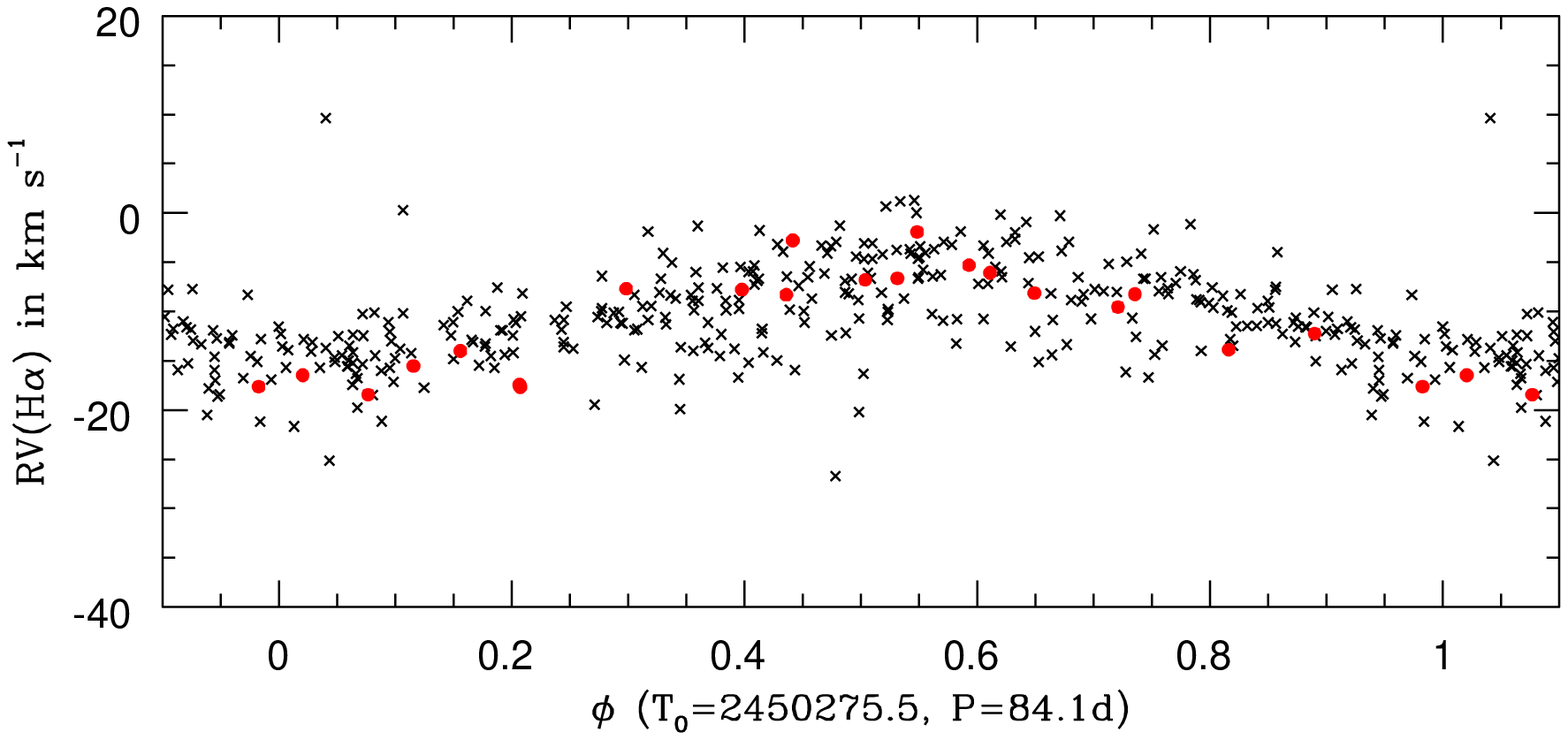}
  \end{center}
\caption{Comparison of first-order moments derived for H$\alpha$ in amateur spectra taken between October 2014 and January 2019 (black crosses) and in TIGRE data (red points). }
\label{rvama}
\end{figure}

\begin{figure*}
  \begin{center}
\includegraphics[width=6cm]{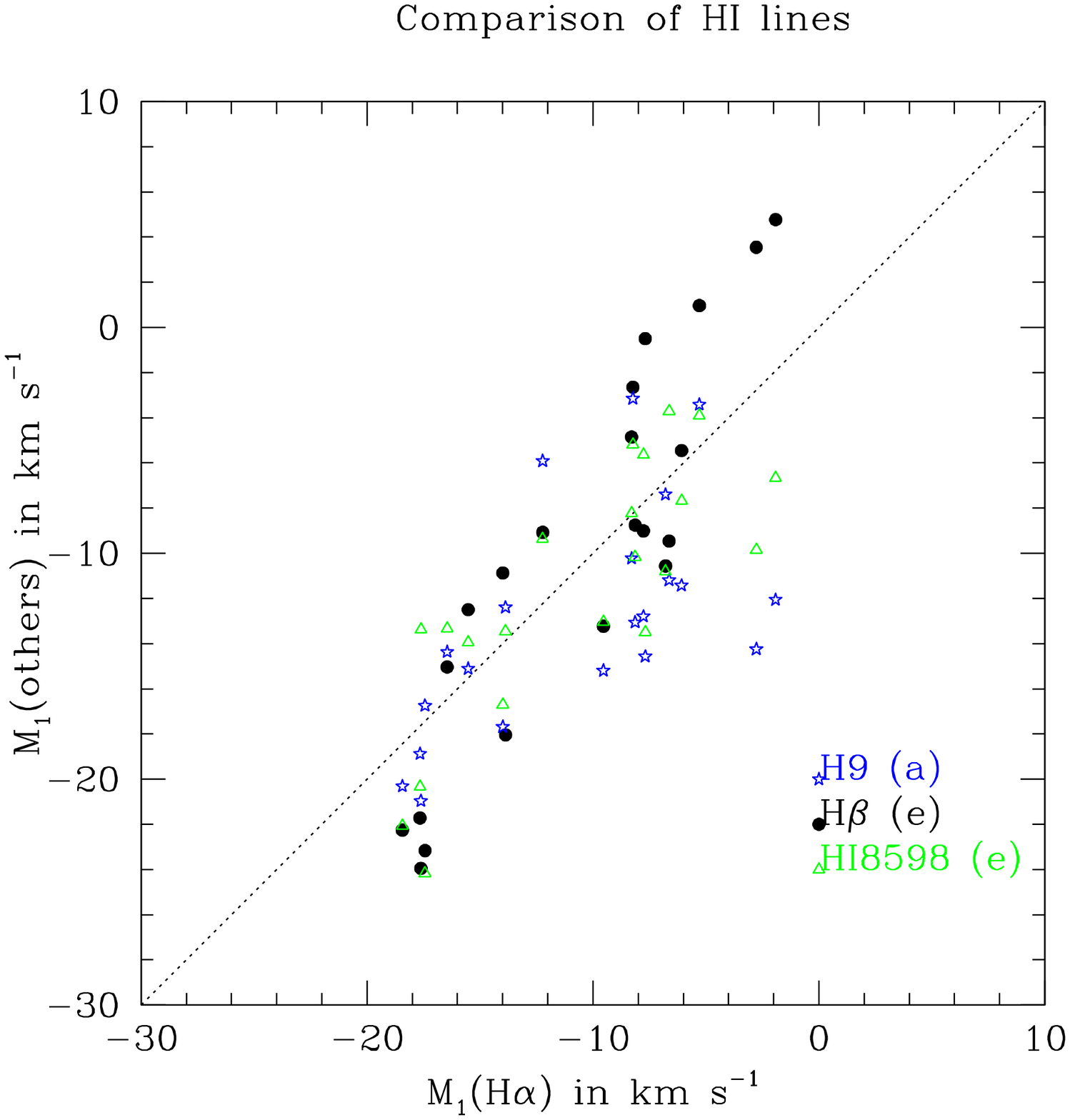}
\includegraphics[width=6cm]{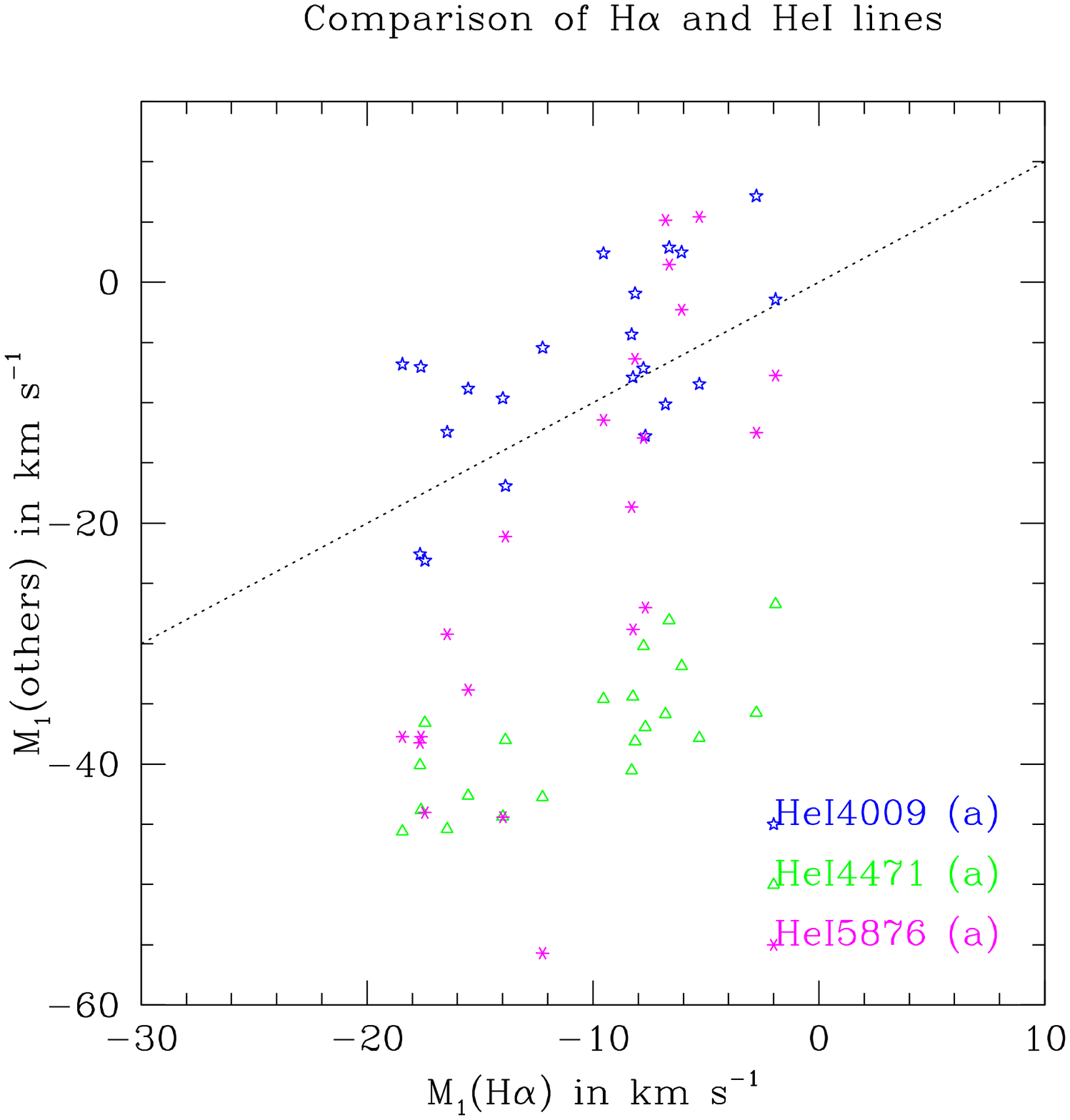}
\includegraphics[width=6cm]{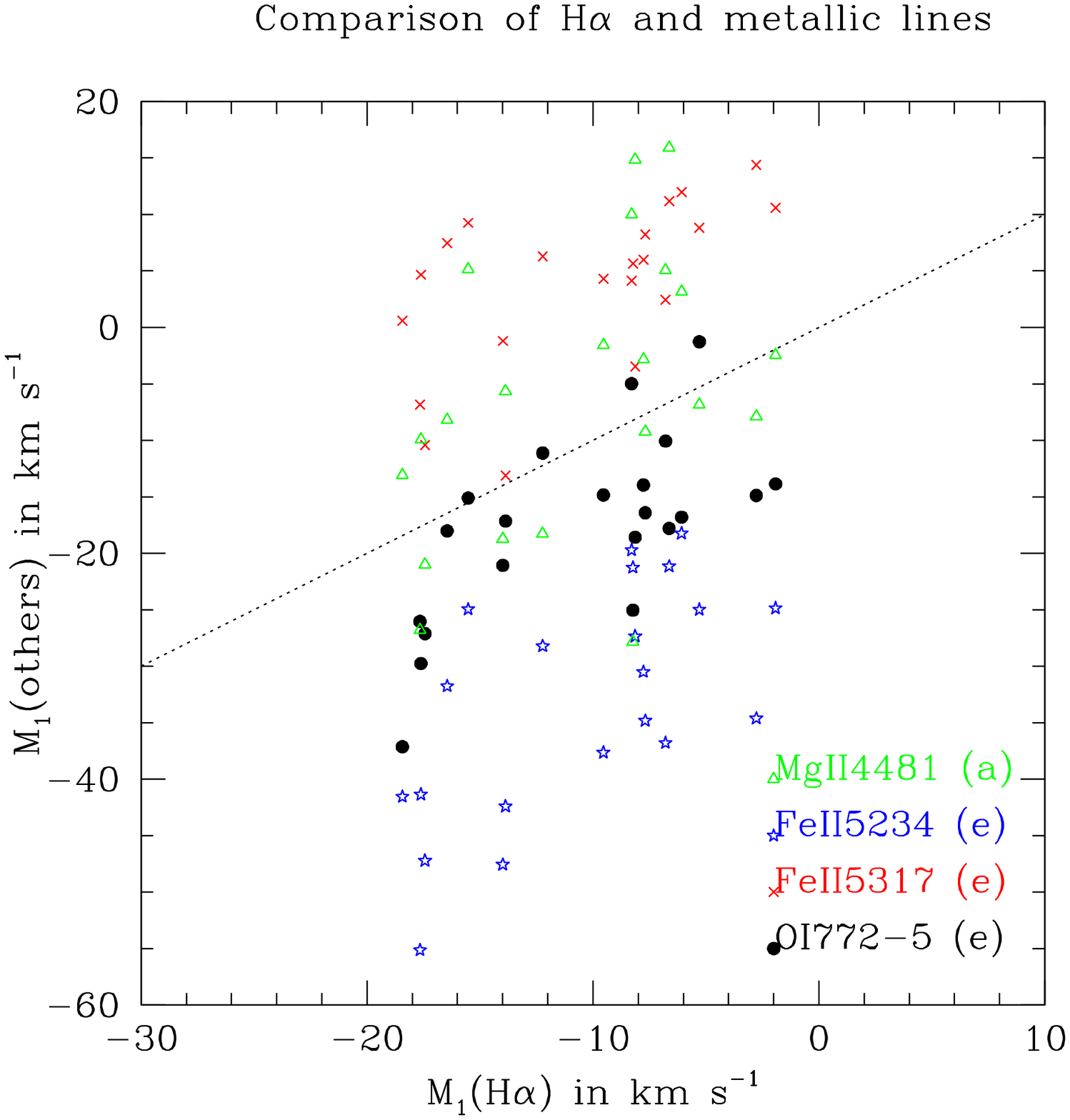}
  \end{center}
\caption{Comparison of first-order moments of H$\alpha$ to those of other lines. }
\label{rvbis}
\end{figure*}

However, the three methods for deriving H$\alpha$ velocities sample very different regions of the disk but yield very similar results. The moment method uses the full line profile, including any potential (direct or indirect) signature of the companion. The two-Gaussian correlation method only examines the line profile in a very small interval near its half-maximum (which occurs near $\pm$245\,\kms), and the mirror method only examines wings, approximately in the intervals between $\pm$375\,\kms\ and $\pm$210\,\kms. These two methods therefore probe regions that are not affected by the companion line because its velocity lies between --100\,\kms\ and 100\,\kms\ \citep{bjo02}, but they could still suffer from the effect of a disk asymmetry triggered by the companion (although the H$\alpha$ emission region is currently much more symmetric, see section 3.2 for more details). Because the regions probed by the three methods are different, we would expect a varying effect on the resulting velocities, hence diverse $RV$s, if the influence of the companion were biasing the results, but this is not the case (Fig. \ref{3rv}). Only a small difference does exist: the first-order moments slightly deviate, preferentially showing less extreme velocities than derived from the other methods (i.e., less negative velocities near minimum velocity and more negative velocities near maximum velocity: black dots are first below the dotted diagonal line and then above it in Fig. \ref{3rv}). If this is a real effect and if it is linked to the direct signature of the secondary (i.e., the low emission recorded by \citealt{bjo02}), then it leads to a change by a few \kms\ at most on both extremes of the RV curve. However, the observed change with respect to the \citet{bjo02} $RV$ curve is not a global reduction in amplitude. The more negative velocities seem to remain unchanged and only the positive velocities appear affected, having been shifted to lower values. Furthermore, as the H$\alpha$ emission strength and the disk asymmetry varied greatly over the years \citep{naz19}, we would expect that the difference, if it is due to the effect of the disk or companion, to change with time, but the opposite appears to be true: a very similar range in velocity is recorded over the years (Fig. \ref{rvama}).  

\begin{figure}
  \begin{center}
\includegraphics[width=8.5cm]{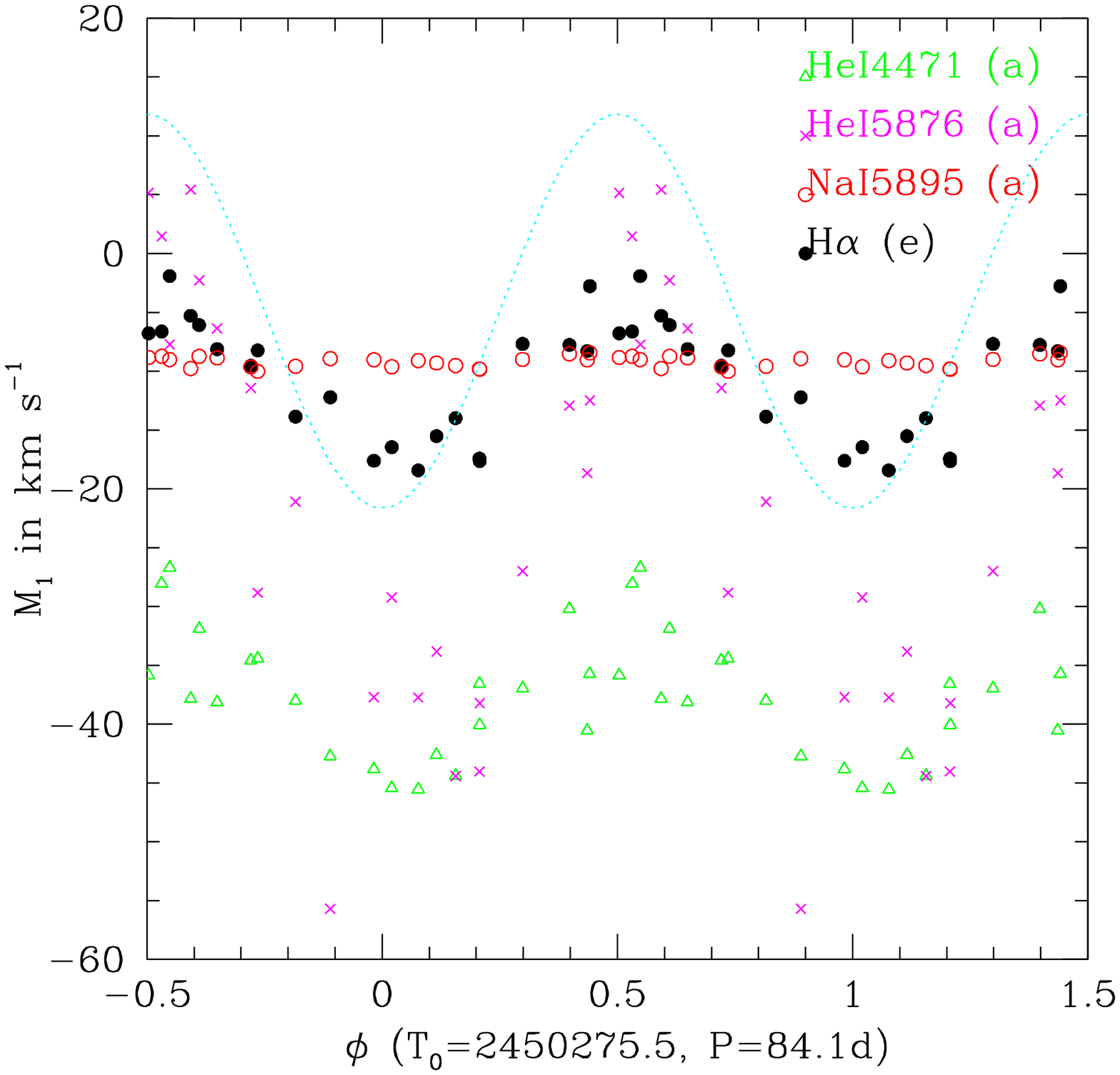}
  \end{center}
\caption{Evolution with phase of first-order moments of H$\alpha$ and other lines. The dotted cyan line provides the orbital solution of \citet{bjo02}. The velocities of a non-varying interstellar line are shown in red to provide a comparison point.}
\label{rvter}
\end{figure}

In Fig. \ref{rvbis} we further compare the first-order moments of H$\alpha$ to those of other lines. A very good correlation is found for other H\,{\sc i} lines regardless of whether they are in absorption or emission, although the blends in H\,{\sc i}\,$\lambda$8437,8502\AA\ lead to a global shift for them. He\,{\sc i} lines remain in absorption, and the velocities of He\,{\sc i}\,$\lambda$4009,4026,4471\AA\ display a neat correlation with those of H$\alpha$, although there is a global shift for the last two lines. He\,{\sc i}\,$\lambda$5876\AA\ appears contaminated by emission in its wings, hence we estimated the moments only in the line core ($v$ in the interval --175 to 175\,\kms). The resulting centroid appears well correlated with the H$\alpha$ velocity, although it covers a wider velocity range. This is best visible in Fig. \ref{rvter} where the velocities are folded with the ephemeris of \citet{bjo02}. Finally, the He\,{\sc i}\,$\lambda$3819,3927,4143\AA\ seem too noisy to allow obtaining good results. The same can be said for many metallic lines, but some of them (Fig. \ref{rvbis}) still appear with velocities similar to H$\alpha$ (although the correlation remains noisy). The difference with the results of Bjorkman et al. therefore remains unexplained.

\begin{figure}
  \begin{center}
\includegraphics[width=8.5cm]{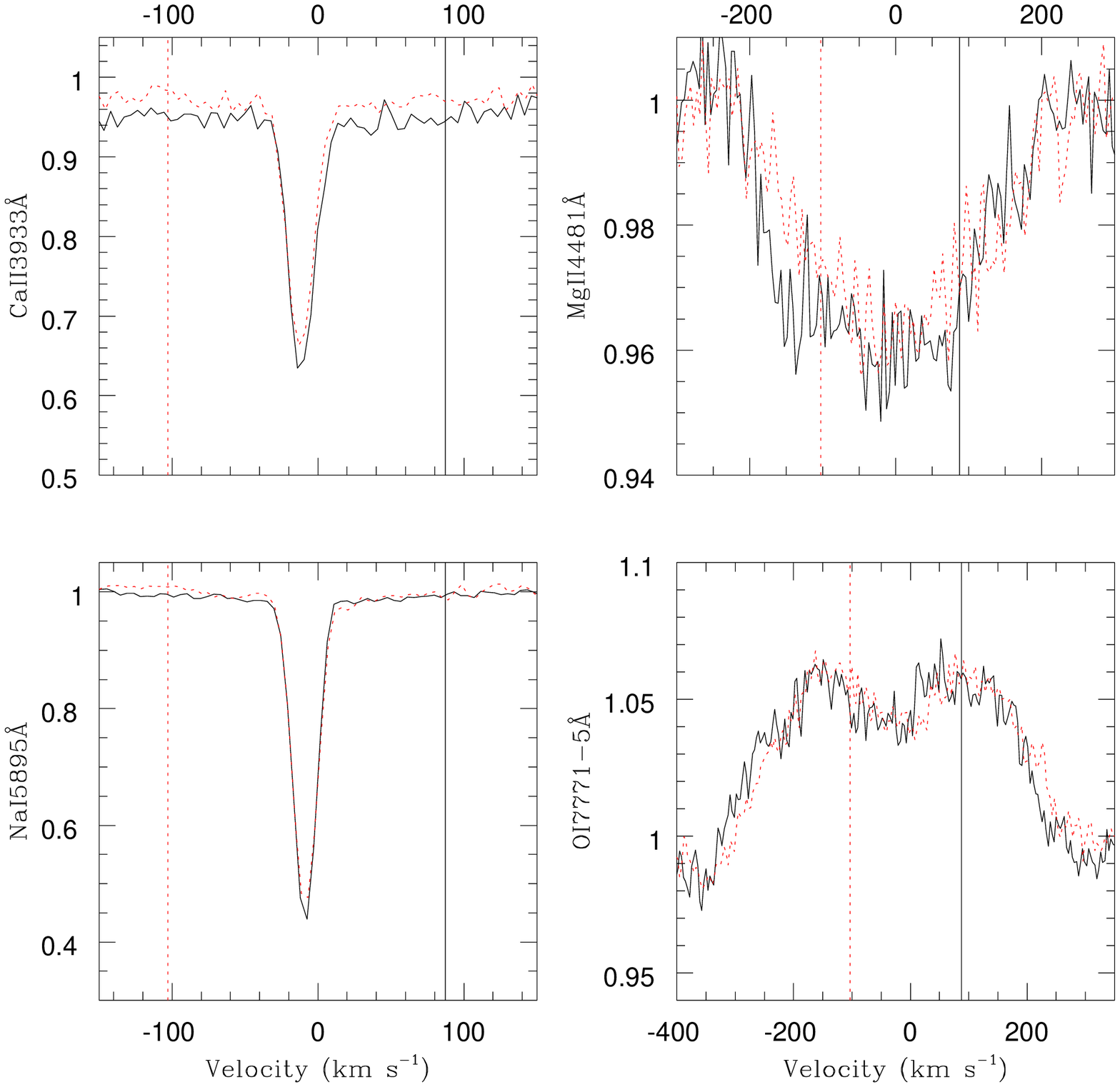}
\includegraphics[width=8.5cm]{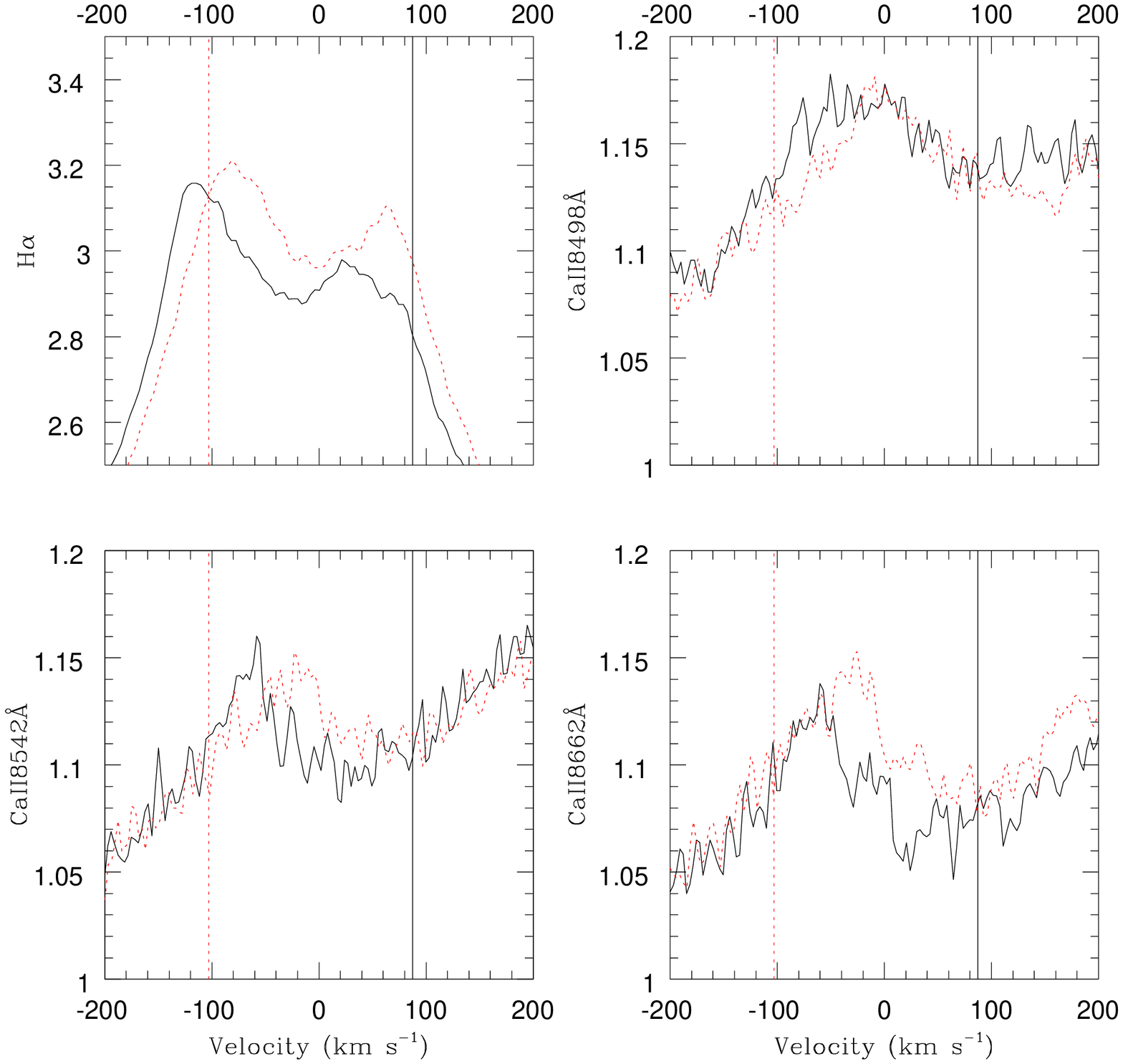}
  \end{center}
\caption{Comparison of spectra taken at two different phases (25 August 2018, $\phi=0.08$, shown as the solid black line, and 29 September 2018, $\phi=0.50$, shown as the red dotted line) around selected lines. The vertical lines indicate the expected companion velocity for the orbital solution of \citet{bjo02}.}
\label{comp}
\end{figure}

\citet{bjo02} had found the direct signature of the companion in the H$\alpha$ line during the quasi-normal star phase of \pa. It was a shallow emission with a full width at half-maximum $FWHM$=200\,\kms\ and an amplitude reaching only 8\% of the continuum. We have examined the TIGRE spectra in quest of lines from the companion. Fig. \ref{comp} compares lines taken at phases corresponding to the two extremes in RVs. While the Be lines display a small shift and the interstellar lines remain stationary, there is no evidence of a component at the expected companion velocity. 

\subsection{Tomography}
\begin{figure}
  \begin{center}
\includegraphics[width=8.5cm]{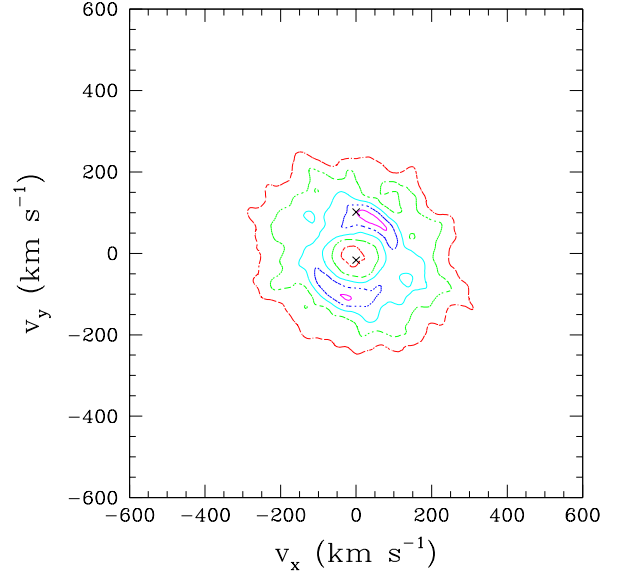}
  \end{center}
\caption{Doppler map for H$\alpha$ in 2018. The magenta, blue, cyan, green, and red contours correspond to amplitudes of 95\%, 80\%, 65\%, 50\%, and 35\% of the maximum intensity. The two crosses indicate the velocities of the secondary (top) and primary (bottom) according to the semi-amplitudes $K$ derived by \citet{bjo02}. The maps correspond to a slice at $v_z=-10$\,km\,s$^{-1}$, the mean value in RV of the system; values of +10 or 0\,km\,s$^{-1}$  provide similar results, however. }
\label{tomo}
\end{figure}

We performed a tomographic analysis of the H$\alpha$ line recorded by TIGRE in the same way as in \citet{naz19}. This technique assumes that the line profile variations are only due to our changing viewing angle because of the orbital motion \citep{hor91}. Therefore, the emitting gas is stationary in the rotating frame of the binary, and each emitting parcel is associated with a specific $(v_x,v_y)$ pair, considering an x-axis pointing from the primary to the secondary and a y-axis in the direction of the secondary motion. At any phase, the radial velocity of an emitting parcel recorded by a terrestrial observer will then be $v(\phi_t)=-v_x \cos(2\pi\phi_t)+v_y\sin(2\pi\phi_t)+v_z$. In this formulation, $\phi_t$ is zero at the conjunction with the secondary in front, so that it is shifted with respect to the orbital phase $\phi$ based on the ephemeris of \citet{bjo02}: $\phi_t=\phi+0.25$. Our Doppler tomography method further uses a Fourier-filtering to suppress some unwanted artifacts \citep{hor91,rau02,rau05}. The resulting Doppler map, shown in Fig. \ref{tomo}, provides the positions of emitting parcels in the {\it \textup{velocity}} space considering that they remain stationary in the rotating frame of the binary. We may note that the tomographic map derived from H$\alpha$ spectra taken by amateurs in 2018 (see \citealt{naz19} for the list) appears very similar to that derived from the TIGRE data alone, showing that the map appearance is not an artifact that is due to a peculiar sampling, for instance.

In this map, the Be disk appears as a large annular region encompassing the stars, with an outer radius slightly larger than 200\,\kms\ and an inner radius $<$50\kms. The maximum intensities (delineated by the dark blue and magenta contours in Fig. \ref{tomo}) occur at a radius of $\sim$100\,\kms, with a radial drop in intensity on either side. An additional drop in azimuth indicates a small asymmetry: the maximum intensities occur near the line joining the two stars and decrease in other directions. Compared to previous years \citep{zha13,naz19}, the disk map in 2018 appears similar to the map recorded in 2016 or 2017. The Doppler map strongly differs, however, with the very asymmetric and thin partial ring of emission observed before 2015 when the H$\alpha$ emission was much weaker.

Tomographic maps of other emission lines were also calculated (H$\beta$, H\,{\sc i}\,$\lambda$\,8545,8598\AA, Fe\,{\sc ii}\,$\lambda$\,5317\AA, O\,{\sc i}\,$\lambda$\,7771--7775\AA). The resulting maps were similar to that of H$\alpha$, although they were noisier because these emissions are weaker.

\section{High-energy emission}
In the X-ray range, \pa\ was first detected during the {\it ROSAT} all-sky survey and then during an \xmm\ slew maneuver, but a detailed spectral description awaited a specific pointing by \xmm\ in mid-November 2013 \citep{naz17}. The latter observation showed that \pa\ exhibits the properties of $\gamma$\,Cas stars: high temperature ($kT\sim$11\,keV), brightness intermediate between ``normal'' B stars and X-ray binaries ($F^{\rm obs}_{\rm X}=1.1\times 10^{-11}$\,erg\,cm$^{-2}$\,s$^{-1}$, $\log[L_{\rm X}/L_{\rm BOL}]\sim-5.5$),  a fluorescent iron line in the iron complex at 6.7\,keV, and variability. In this section, we derive the X-ray properties exhibited by \pa\ in 2018, and examine them in light of the simultaneous optical monitoring. We also compare them to the older X-ray data. 

\begin{figure}
  \begin{center}
\includegraphics[height=8.5cm,angle=-90]{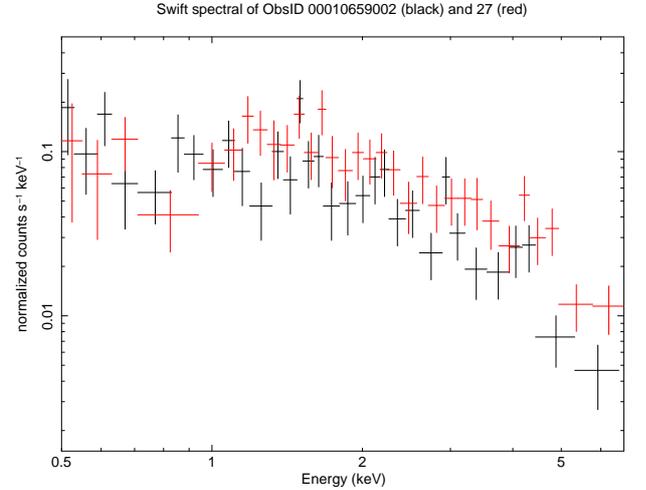}
  \end{center}
\caption{ Two X-ray spectra of \pa\ recorded by \sw, showing the intense emission at high energies at all times, but also clear variations. }
\label{specx}
\end{figure}

\subsection{Spectral fitting}
Count rates (Table~\ref{journal}) already provide some information on the X-ray properties of \pa, notably indicating significant variability (constancy is rejected with a significance level well below 0.1\%). However, to study the system in more detail, the X-ray spectral distribution should be analyzed (Fig. \ref{specx}). To this aim, we used Xspec v12.9.1p and considered reference solar abundances from \citet{asp09}. A visual examination of the spectra reveals that despite precautions (see Section 2.2), the spectra appear somewhat erratic or very noisy below 0.5\,keV. These spectral bins were therefore not considered in our spectral fitting. As is usual for massive stars, we used an absorbed optically thin thermal emission model of the type $tbabs\times phabs \times apec$. The first absorption component represents the interstellar contribution, fixed to $3.6\times10^{20}$\,cm$^{-2}$ \citep{naz18}, while the second accounts for additional local absorption. A single thermal component was used because this is sufficient to reach a good fit. \citet{naz17} added a Gaussian line to account for the fluorescent iron line, but the low signal-to-noise ratio of \sw\ spectra (compared to \xmm\ data) renders this addition unnecessary.

We performed several fitting trials. First, we simultaneously fit all 30 \sw\ spectra with the same model. The reduced $\chi^2$ of the best fit reached a rather high value (1.80), which could be expected in view of the significant count rate changes from one observation to the next. We then kept the same temperature and absorbing column for all spectra, but allowed for different normalization factors. At 1.24, the reduced $\chi^2$ was much improved. The best-fit local column was (5.6$\pm$0.5)$\times10^{22}$\,cm$^{-2}$ and the best-fit temperature 16.5\,keV, with a 68\% (1$\sigma$) confidence interval of 13.2--21.0\,keV. A last simultaneous fitting considered only the temperature as common to all data. It yielded an even better reduced $\chi^2$ (0.93), with a best-fit temperature of 14.8\,keV (with a 68\% confidence interval of 12.2--18.4\,keV). 

\begin{figure}
  \begin{center}
\includegraphics[width=8.5cm]{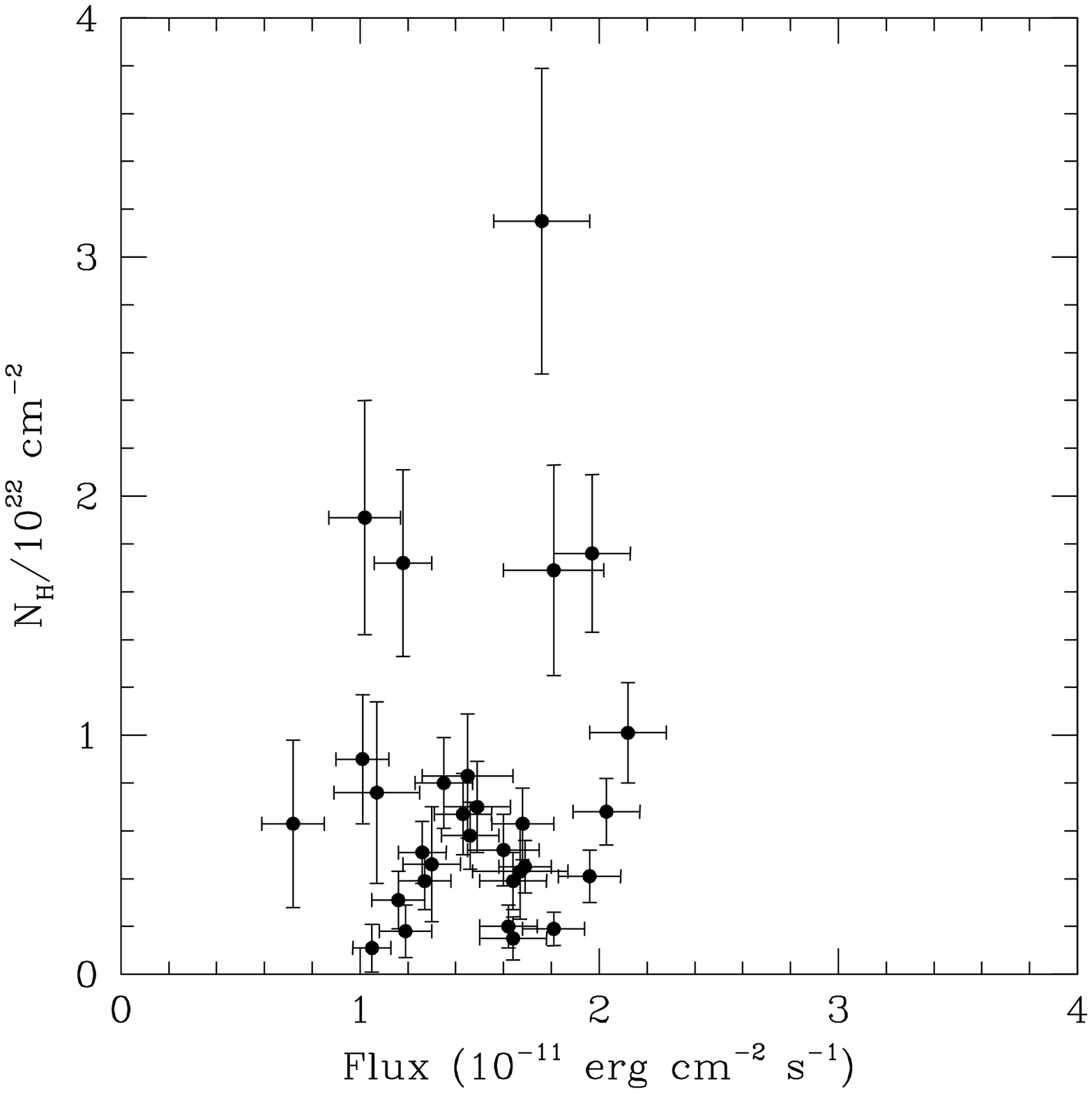}
  \end{center}
\caption{Comparison between the local absorbing columns and the observed fluxes. }
\label{nhfx}
\end{figure}

We then performed individual fittings, with local absorbing column, temperature, and normalization factor as free parameters. However, the fits often meandered because of the low signal-to-noise ratio, resulting in unrealistic large error bars on the derived fluxes. We therefore decided to fix the temperature to 14.8\,keV, as found in the simultaneous fitting, and found that the fit quality remained very similar (with a small modification of the second decimal of the reduced $\chi^2$ at most). Table~\ref{journal} reports the results of this fitting procedure. When the temperature was instead fixed to 11.0\,keV, as found from \xmm\ data, the results were only marginally affected, with a second decimal change of the reduced $\chi^2$, for example. As a check, we compared the observed fluxes derived from this constrained fitting with the count rates (i.e., values derived without any fitting), and found a good correlation between them ($\rho$=70\%). As a second check, we also fit spectra obtained from the online tool (see footnote 1 in section 2.2), and their results agree well with those derived from locally processed spectra. We therefore report in Table~\ref{journal} only the latter.

Following recommendations of the XRT team for such bright sources \citep[see also][]{naz18zp}, the spectral fitting was also performed considering the possibility of energy scale offsets (command ``gain fit'' under Xspec, with slope fixed to 1). However, not only did the spectral parameters agree with those found from the usual fitting, but (1) the $\chi^2$ improvement was marginal, (2) very different offset values provided very similar statistics, and (3) offset values appeared not significant because they agreed with zero within the errors (or twice these 1$\sigma$ errors). Because this adjustment provided no significant improvement on the fits, we discuss only the fit achieved without these offsets here.

\begin{figure*}
  \begin{center}
\includegraphics[width=6.cm]{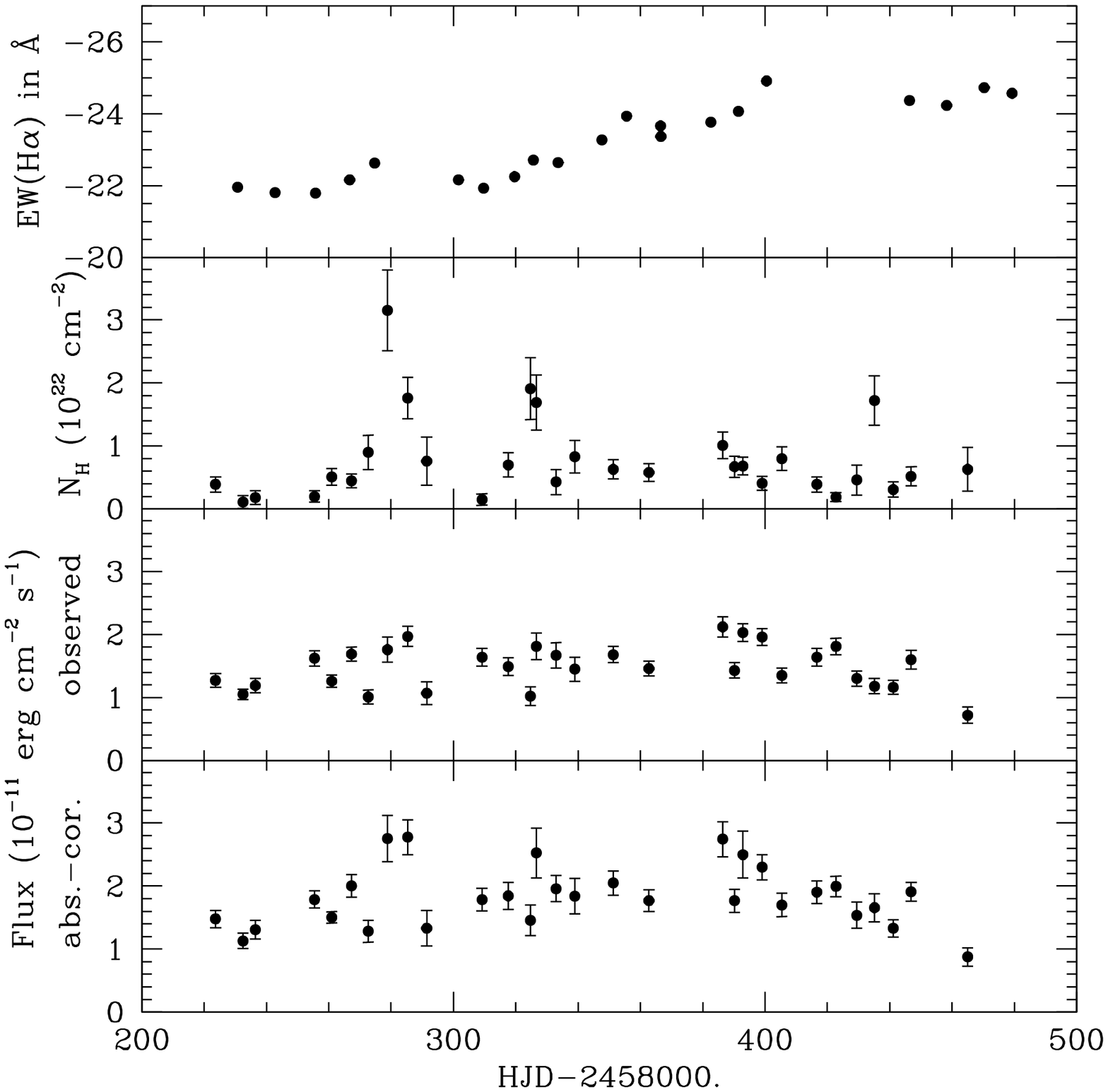}
\includegraphics[width=6.cm]{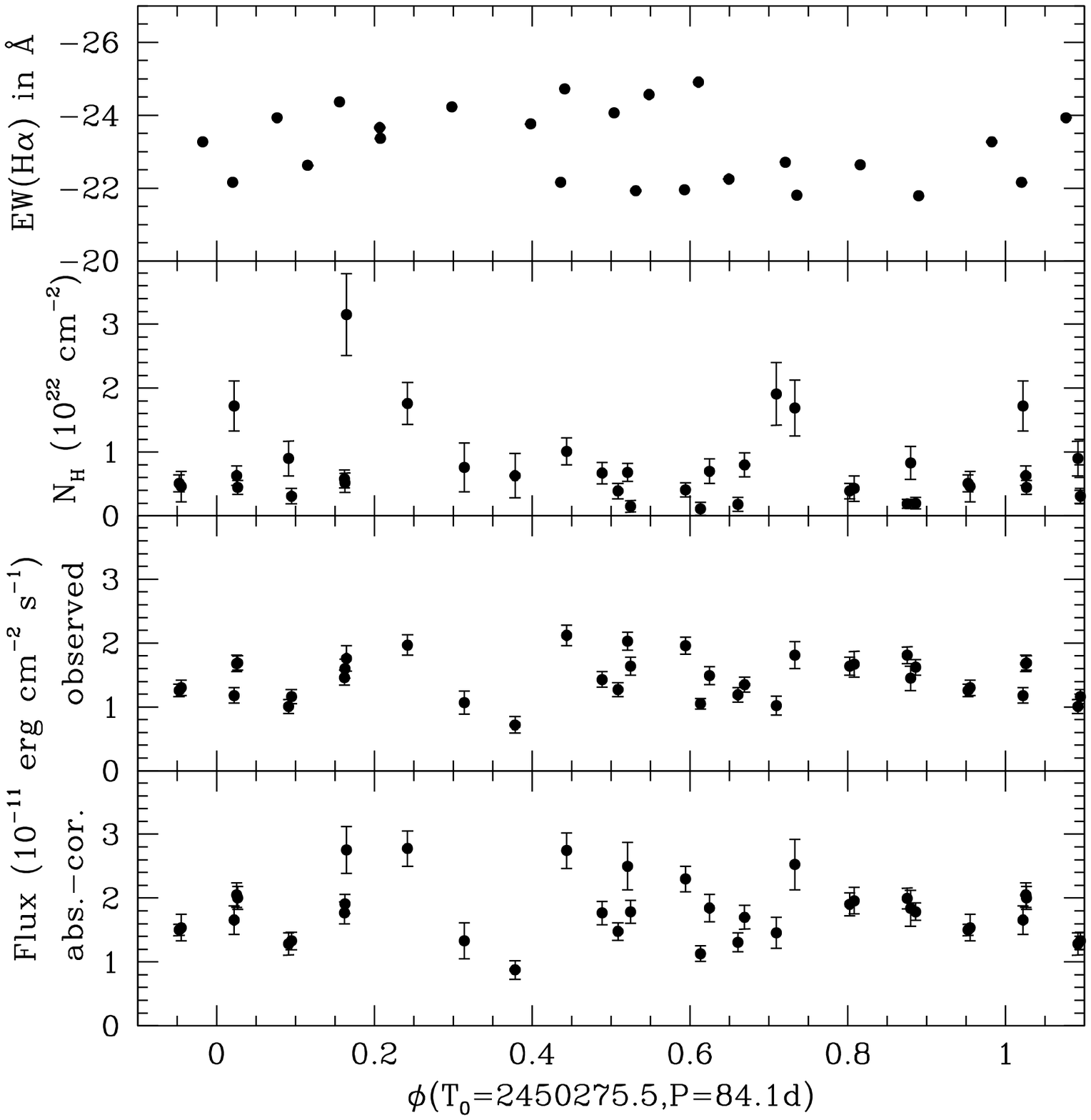}
\includegraphics[width=6.cm]{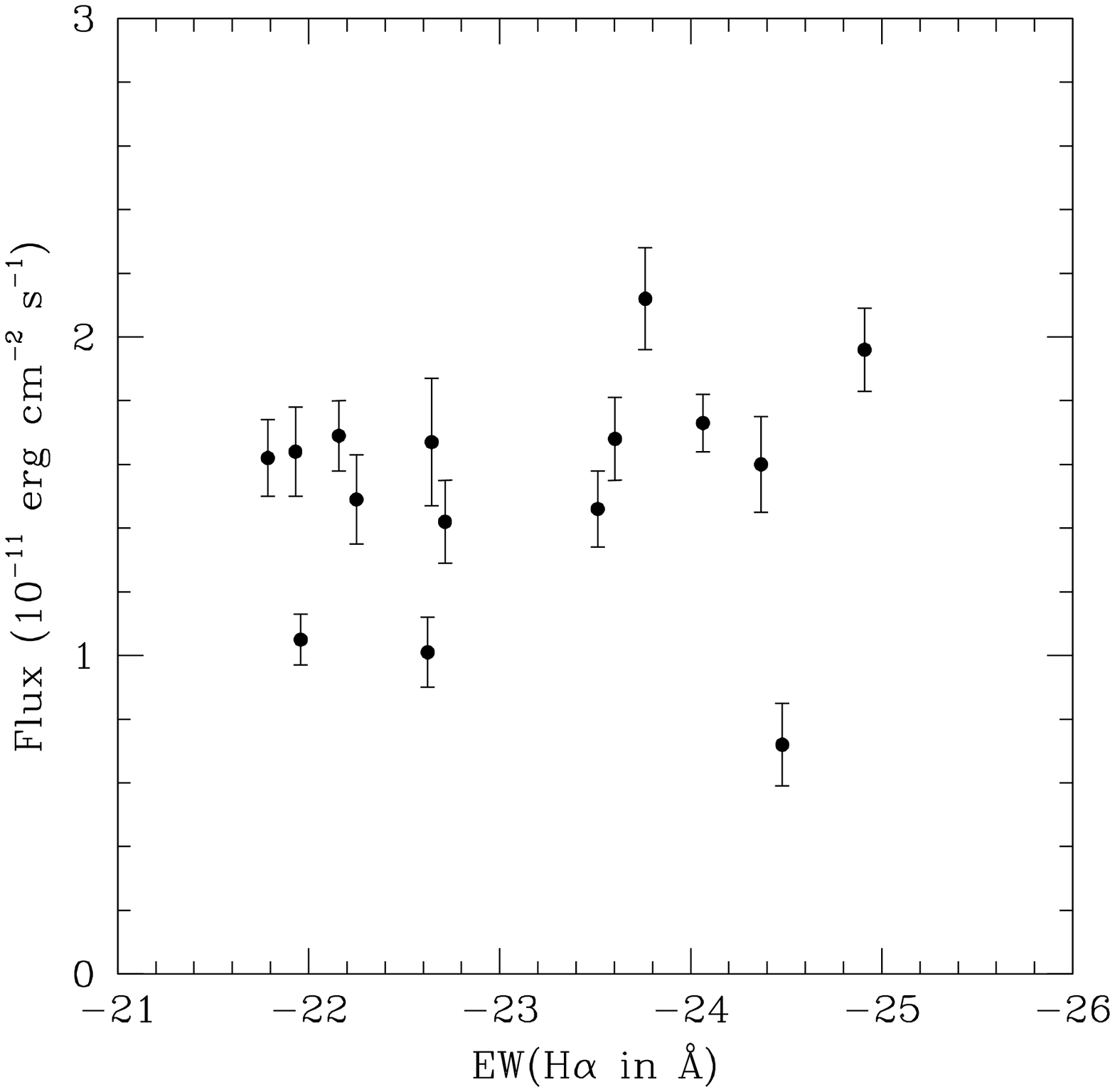}
  \end{center}
\caption{Evolution with time (left) or orbital phase (middle, using an average ephemeris from \citealt{bjo02}) of the X-ray fluxes, best-fit local absorbing columns, and $EW$s of H$\alpha$ in 2018 for \pa. The right panel directly compares the $EW$ of H$\alpha$ measured on TIGRE spectra and the observed X-ray flux of the closest \sw\ exposure(s). }
\label{sw}
\end{figure*}

\subsection{Behavior in 2018}
During the 2018 campaign, the observed X-ray flux of \pa\ varied from 0.7 to 2.1$\times 10^{-11}$\,erg\,cm$^{-2}$\,s$^{-1}$ (Table~\ref{journal}). Normalization factors varied by a larger factor (3.2 versus 2.9) because the absorbing column also changed. However, the low-energy bins have low signal-to-noise ratios and are therefore less reliable: while the local column seems to change (see previous section), its error bar remains large and caution should be taken when interpreting it. We may nevertheless note the presence of possible high-absorption events. This is not unusual for $\gamma$\,Cas stars \citep{ham16}, but they generally corresponded to times of lower (observed) X-ray fluxes. In our data, no strong correlation exists between absorbing column and fluxes: the correlation coefficient $\rho$ is positive but only amounts to 12\% when columns are compared to the observed fluxes and 46\% when the noisier absorption-corrected fluxes are considered (Fig. \ref{nhfx}). 

Figure \ref{sw} further shows the evolution of the X-ray fluxes and local absorbing columns with time and orbital phase: no coherent behavior is detected. Furthermore, in the same figure, the flux variations are compared with the evolution of the strength of the H$\alpha$ line, which is a diagnostics of the disk properties. During 2018, the line strength slowly increased with time, the $EW$ passed from --22 to --24.5\AA. In contrast, the X-ray fluxes do not display any obvious trend with time, neither does the local absorbing column: both rather seem to fluctuate continuously. The right panel of Fig. \ref{sw} directly compares X-ray fluxes and $EW$s when the X-ray exposures are associated with the closest optical spectra (see the last column of Table \ref{journalHRT}; averages were used when two X-ray datasets corresponded to one optical spectrum or vice versa). The figure shows large scatter. The correlation coefficient between these two parameters is only 15\%, which demonstrates the absence of a strong and direct correlation between X-ray flux and H$\alpha$ emission strength.  Regardless of the time, the range in flux values remains rather constant. Similarly, no coherent behavior is found with phase for  optical and X-ray diagnostics, despite a rather high disk inclination (50--75$^{\circ}$, with 70$^{\circ}$ favored, see \citealt{bjo02}). The absence of (strong) orbital effects underlines that H$\alpha$ and X-ray emission processes are independent from the presence of a companion. It may further be related to the fact that the disk now lacks the strong asymmetry noted in earlier data (see section 3.2 and \citealt{naz19}). 

\subsection{Long-term behavior}
Examining the behavior of \pa\ on longer timescales implies a comparison with previous X-ray observations that were taken by {\it ROSAT} and \xmm, although only the pointed \xmm\ exposure (taken on $HJD=2\,456\,614.422$) provided spectral information \citep{naz17,naz18}. The best-fit temperature found for \sw\ spectra is only slightly higher than the temperature found for the pointed \xmm\ observation (14.8 or 16.5\,keV versus $\sim$11.0\,keV). Considering the error bars, this difference is marginal: it is only at the 2$\sigma$ level. In addition, the fit quality does not significantly change when the \xmm\ temperature is used to fit the \sw\ data. The available data therefore do not suggest strong temperature variations in the hot plasma of \pa.

The local absorbing columns found in individual fits vary between $10^{20}$ and $3\times10^{22}$\,cm$^{-2}$, a wide range that encompasses the \xmm\ value ($2\times10^{21}$\,cm$^{-2}$). However, the \sw\ observations are 1\,ks snapshots, while the \xmm\ exposure was 55\,ks long. During this exposure, the source brightness varied by a factor 2--3 over a few hours (see Fig. 2 of \citealt{naz17}), a similar range as observed in \sw\ observations over several months. It therefore appears meaningful to compare the column found from the fitting the whole \xmm\ exposure with that derived from the simultaneous fitting of all \sw\ spectra ($5.6\times10^{22}$\,cm$^{-2}$, see above). The local absorption is then found to be significantly (7$\sigma$) higher in \sw\ observations. Because no systematic difference between \sw\ and \xmm\ (i.e., due to imperfect cross-calibration) has been reported, this absorption difference is most probably real. 

In the \sw\ data, \pa\ displays an average flux of $F^{\rm obs}_{\rm X}=(1.38\pm0.04)\times 10^{-11}$\,erg\,cm$^{-2}$\,s$^{-1}$. In the pointed \xmm\ exposure, \pa\ showed $F^{\rm obs}_{\rm X}=(1.055\pm0.004)\times 10^{-11}$\,erg\,cm$^{-2}$\,s$^{-1}$ \citep{naz18}, therefore the X-ray level of \pa\ appears somewhat higher in 2018 than in 2013. \pa\ was also detected by \xmm\ on $HJD=2\,453\,145.919$ during slew maneuvers. It is reported in the slew survey clean catalog, v2.0, as XMMSL2 J222517.0+012226 with an EPIC-pn count rate of $2.8\pm1.1$\,cts\,s$^{-1}$ in 0.2--12.\,keV. In the same energy band, the 3XMM-DR8 catalog reports the \pa\ detection from the pointed observation as 3XMM J222516.6+012238 with an EPIC-pn\footnote{\citet{naz17} incorrectly assumed that the slew count rates refer to all EPIC and not to the EPIC-pn camera alone.} count rate of $2.242\pm0.008$\,cts\,s$^{-1}$. The slew count rate is therefore formally 25\% higher than that in the pointed observation, but it is indistinguishable in view of its large error bar. \citet{naz17} also compared the {\it ROSAT} detection and the pointed \xmm\ observation. The {\it ROSAT} count rate formally indicates a flux in the 0.1--2.0\,keV energy band that is $\sim$30\% lower, but the values were compatible at 3$\sigma$ considering the rather large {\it ROSAT} error.

Overall, the average X-ray level of \pa\ thus seems to vary in a moderate way (up to 40\%) over time. However, the short-term variations are of a large amplitude: a factor of 3 during the 2018 \sw\ monitoring, and a factor of 2--3 during the 55\,ks \xmm\ exposure. A random snapshot of \pa, such as the \xmm\ slew observation or any individual \sw\ exposure, therefore has a high probability of deviating from the average by a non-negligible amount. When we take these short-term variations into account and consider the similar average levels, the only possible conclusion is that the X-ray emission level of \pa\ has not significantly changed over time.

In parallel, the disk of \pa\ has undergone dramatic changes, as traced by its H$\alpha$ line. During the \sw\ monitoring, the H$\alpha$ line was very strong ($EW\sim-23$\,\AA) and the disk appeared quite symmetric (see section 3.2). In contrast, at the time of the pointed \xmm\ observation, the H$\alpha$ emission was much weaker, and the photospheric absorption was clearly detectable \citep{naz19}. The closest amateur spectrum, taken only one day later, has $EW=-1.73$\AA\ \citep{naz19}, with $EW$ estimated in the same way as in previous section. Furthermore, the disk was then much more asymmetric and much less extended \citep{naz19}. In this context, it is also important to note that ASAS-SN photometry indicates a simultaneous overall brightening of \pa\ by $\sim$0.4\,mag in $V$ band (Fig. \ref{photom}).

Unfortunately, much less information is available for the time of the other X-ray observations. The {\it ROSAT} survey observations took place in 1990--1991, when \pa\ was rapidly transitioning from an active to a quiet phase \citep{bjo02}: very precise observing dates would be needed for a meaningful statement, but they are not available. The $EW$ of the H$\alpha$ line at the time of the \xmm\ slew observation ($HJD=2\,453\,145.919$) is unknown: the closest spectrum, taken three months later, is reported by \citet{zha13} with a moderate $EW$ of --3.53\AA.

To further specify the extent of the disk variation between 2013 and 2018, we now try to quantitatively assess the disk size. The separation between the violet and red peaks of the H$\alpha$ emission profile can be related to the radius of the emission zone if the disk rotation is Keplerian \citep{zam19}. When we use the observed separation ($\sim$320\,\kms\ in 2013 and $\sim$140\,\kms\ in 2018, see \citealt{naz19}) as well as the projected rotational velocity of the star ($v \sin i\sim250$\,\kms) and stellar radius ($R_*=6.1R_{\odot}$) from \citet{bjo02}, Eq. 2 of \citet{zam19} yields a radius for the emission region of 2.4\,$R_*$=14.9\,$R_{\odot}$ in 2013 and 12.8\,$R_*$=77.8\,$R_{\odot}$ in 2018, that is, an increase by a factor of 5 in the extension of the H$\alpha$ emission region. The separation between the Be star and its companion was determined to be $a \sin i=0.96$\,AU=207\,$R_{\odot}$, or 220\,$R_{\odot}$ for the favored inclination, 70$^{\circ}$ \citep{bjo02}. This means that the orbital separation was 14.8 times the disk radius (as determined from the H$\alpha$ emission) in 2013, but only 2.8 times that radius in 2018. In section 3.1 we found that the velocity amplitude of the primary in 2018 was half the value reported by \citet{bjo02}, which implies a separation $a$ smaller by 7\%. For the same inclination, this separation would correspond to 13.7\,$R_{disk}$ in 2013 or 2.6\,$R_{disk}$ in 2018. Because the absorbing column appears larger when the disk emission is stronger, this may readily be interpreted in terms of more material being available along the line of sight to emit as well as to absorb X-rays (the latter being observed while the former is not).

How does \pa\ compare with other $\gamma$\,Cas stars? Unfortunately, only two such stars were simultaneously monitored in X-ray and optical domains. The first is $\gamma$\,Cas itself. It simultaneously brightens at optical and X-ray wavelengths, as demonstrated by the clear correlation between the X-ray fluxes and the V-band magnitudes found by \citet{rob02} and \citet{mot15}, apparently without delay, and we recall that \citet{smi12} found that $\gamma$\,Cas brightens in the optical as its disk emission strengthens. Furthermore, it also exhibited an increased local absorption of X-rays when its H$\alpha$ emission was stronger \citep{smi12}. The second example is HD\,45314. This star was monitored during large variations of its disk by \citet{rau18}: the strong H$\alpha$ weakening ($EW$ passing from --22.6 to --8.5\AA), although not as extreme as for \pa, corresponded to a flux decrease in X-rays of an order of magnitude ($F^{\rm obs}_{\rm X}$ changing from 11. to 1.2$\times 10^{-13}$\,erg\,cm$^{-2}$\,s$^{-1}$). This was interpreted as reflecting the fact that much less disk material was then available for X-ray emission. No obvious relationship was found between local absorbing column and X-ray flux or $EW(H\alpha)$, however, but the spectral model was different in the minimum flux stage because the thermal emission was also much softer at the time, rendering a direct comparison of the absorption of the hard component more awkward. While the variation in absorption seems similar to that recorded in $\gamma$\,Cas, the behavior of \pa\ therefore surprisingly departs from that of its fellow $\gamma$\,Cas stars by lacking the large change in X-ray flux observed for both $\gamma$\,Cas and HD\,45314 in reaction to the disk variations (traced by H$\alpha$ emission and/or V-band magnitudes). 

Theoretically, both scenarios under consideration for explaining the $\gamma$\,Cas phenomenon (accretion and star-disk interactions) expect changes in the disk to lead to changes in X-ray emission level. This is quite obvious for the star-disk interaction scenario because the X-ray emission directly arises in or close to the disk and is therefore expected to depend sensitively on its properties. However, the emission is assumed to take place close to the star in this model, therefore the disk density in the inner regions could be more important than the disk extent itself. 

On the other hand, scenarios involving accretion onto a compact object could be more sensitive to variations in the outer regions of the disk. Be X-ray binaries display changes in their X-ray emission when the disk varies (e.g., type II outbursts, see \citealt{rei11}). Some delay may be involved in this case, however. For accretion onto a neutron star in the propeller stage, \citet{pos17} estimated a delay of $\sqrt{R_L^3/2GM_{NS}}$ (where $R_L$ is the Roche-lobe radius of the neutron star) for free-fall time inside the neutron star's Roche lobe. Considering Kepler's third law $4 \pi^2 a^3 = P^2 G (M_{Be}+M_{NS})$ and $R_L\sim0.5 a M_{NS}/(M_{Be}+M_{NS})$ (eq. 17 in \citealt{pos17}) yields a delay $\sim P/8\pi$ or 3\,d for \pa. To this should be added the time for any stellar mass-loss or outer disk material to flow to the Roche lobe of the neutron star. However, because we focus here on the slow and large H$\alpha$ variations, which took years (see Fig. \ref{photom}), this delay is negligible when we compare the 2013 and 2018 epochs. In this context, it may be useful to note that for the mass ratio of 0.16 determined by \citet{bjo02}, the Roche-lobe radius of the primary reaches 54\% of the separation, or 120$R_{\odot}$, so that no overflow seems possible even in 2018\footnote{If we were to keep the secondary velocity amplitude of \citet{bjo02} but considered half the value for the primary amplitude (see section 3.1), the mass ratio would then be halved and the Roche-lobe radius of the primary increased to 150$R_{\odot}$.}. Disk Roche-lobe overflow therefore seems prohibited at all times. Instead, we might imagine that the stellar wind is responsible for accretion instead of the disk material. However, a strong decoupling between disk (as traced by the H$\alpha$ line in this scenario) and wind (as traced by X-rays in this scenario) is then required in \pa\ {\it \textup{but}} not in the other two $\gamma$\,Cas stars to explain observations. Finally, we may note that orbital modulation of the X-ray emission appears widespread for systems with accreting white dwarfs \citep[e.g.,][]{szk96,pat98,par05}, although none is observed here.

\section{Conclusions}

We conducted a monitoring of \pa\ in 2018 in the optical and X-ray domains.
At the time, \pa\ appeared very active, reaching emission levels that had not been seen since 1990 \citep{naz19}. The H$\alpha$ line shows clear velocity variations, which can be phased with the ephemeris of \citet{bjo02}. However, the positive velocity values recorded two decades ago (when the H$\alpha$ line was in absorption) are not seen during our 2018 monitoring (with the same line being in emission). As a consequence, the velocity amplitude has been halved. This does not seem to be linked to a bias introduced by the companion emission line or its influence on the disk structure.

X-ray variations by a factor of 3 are recorded during our monitoring. However, they show no obvious link with orbital phase or with H$\alpha$ line strength. They most probably reflect the short-term X-ray variability seen in $\gamma$\,Cas stars. Comparing the properties to older X-ray detections, we find that the overall X-ray flux level of \pa\ remains stable, despite an increase in optical brightness (V magnitude) and large changes in disk structure and extent (as traced by the H$\alpha$ line). An increase in absorbing column is potentially detected, however.

This stability appears at odds with observations of other $\gamma$\,Cas stars, which showed clear correlations between their X-ray and optical ($EW(H\alpha)$ or V-magnitude) emissions, and also possibly with scenarios proposed to explain the $\gamma$\,Cas behavior. The surprising behavior of \pa\ therefore points toward a missing ingredient in the considered scenarios. Unfortunately, it is quite difficult to point out the exact reason for this discrepancy because the number of $\gamma$\,Cas stars that are monitored in optical and X-rays is quite limited for the moment. Such monitoring should therefore be performed for more stars to assess exactly whether \pa\ is truly an exception or if a range of behaviors are actually present in $\gamma$\,Cas stars, with \pa\ and $\gamma$\,Cas/HD\,45314 at the extremes. Extending the wavelength coverage to the NIR and UV also seems relevant to further assess the wind and circumstellar material as well as the properties of the companion.

\begin{acknowledgements}
We thank C. Motch, R. Lopes de Oliveira, and A. Miroshnichenko for fruitful discussions. Y.N. and G.R. acknowledge support from the Fonds National de la Recherche Scientifique (Belgium), the Communaut\'e Fran\c caise de Belgique, the European Space Agency (ESA) and the Belgian Federal Science Policy Office (BELSPO) in the framework of the PRODEX Programme (contract XMaS). We are grateful to the \sw\ team, especially Kim Page, for their help and good advice. ADS and CDS were used for preparing this document. 
\end{acknowledgements}

\end{document}